\documentclass[useAMS,usenatbib]{mn2e}

\usepackage{graphicx}
\usepackage{amsmath, amssymb}
\usepackage{float}
\usepackage{subfigure}
\usepackage{color}
\usepackage{xspace}
\usepackage{url}
\usepackage{hyperref}

\newcommand{\gaia}{\textit{Gaia}\xspace}

\numberwithin{equation}{section}
\title[\gaia transient detection efficiency: hunting for nuclear transients]{ \gaia transient detection efficiency: hunting for nuclear transients}
\author[N. Blagorodnova et. al. ] {N. Blagorodnova$^{1}$\thanks{email:nblago@ast.cam.ac.uk}
 S. Van Velzen$^{2}$\thanks{Hubble Fellow;email:sjoert@jhu.edu}, D. L. Harrison,$^{1,3}$  S. Koposov,$^{1,4}$  
   \newauthor  S. Mattila,$^{5,6}$ H. Campbell,$^{1}$  N. A. Walton$^{1}$  and \L . Wyrzykowski$^{7}$ \\
$^{1}$Institute of Astronomy, University of Cambridge, Madingley Road, Cambridge, CB3 0HA, UK\\
$^{2}$The Johns Hopkins University, Baltimore, MD 21218, USA\\
$^{3}$Kavli Institute for Cosmology, University of Cambridge, Madingley Road, Cambridge, CB3 0HA, UK\\
 $^4${Moscow MV Lomonosov State University, Sternberg Astronomical Institute, Moscow, 119992, Russia}\\
 $^{5}$ Finnish Centre for Astronomy with ESO (FINCA), University of Turku, V\"ais\"al\"antie 20, FI-21500 Piikki\"o, Finland \\
 $^{6}$  Tuorla Observatory, Department of Physics and Astronomy, University of Turku, V\"ais\"al\"antie 20, FI-21500, Piikki\"o, Finland\\
$^7$Warsaw University Observatory, Al. Ujazdowskie 4, 00-478 Warszawa, Poland
}

\begin{document}

\date{Accepted 2015 October 2. Received 2015 August 6; in original form 2015 April 29}

\pagerange{\pageref{firstpage}--\pageref{lastpage}} \pubyear{2015}

\maketitle

\label{firstpage}

\begin{abstract}

We present a study of the detectability of transient events associated with galaxies for the \gaia European Space Agency astrometric mission. We simulated the on-board detections, and on-ground processing for a mock galaxy catalogue to establish the properties required for the discovery of transient events by \gaia, specifically tidal disruption events (TDEs) and supernovae (SNe). Transients may either be discovered by the on-board detection of a new source or by the brightening of a previously known source. We show that \gaia transients can be identified as new detections on-board for offsets from the host galaxy nucleus of 0.1--0.5\,arcsec, depending on magnitude and scanning angle. The \gaia detection system shows no significant loss of SNe at close radial distances to the nucleus. We used the detection efficiencies to predict the number of transients events discovered by \gaia. For a limiting magnitude of 19, we expect around 1300 SNe per year: 65\% SN Ia, 28\% SN II and 7\% SN Ibc, and $\sim$20 TDEs per year.

\end{abstract}


\section{Introduction}

 \gaia is an European Space Agency (ESA) satellite mission, successfully launched in 2013 December.  It was inserted into a Lissajous orbit around the L2 Lagrange point of the Sun-Earth system. \gaia is an astrometric mission \citep{Perryman2001} which will improve upon the previous ESA  astrometry mission, {\it Hipparcos},  with 10 000 times the number of stars observed and an increase in the parallax and proper-motion accuracy attained by 2 orders of magnitude, \citep{vanLeeuwen07}. The \gaia catalogue will amount to about one billion stars, or 1 per cent of the Galactic stellar population, complete to 20$^{th}$ magnitude. This catalogue will consist of positions, proper motions, parallaxes, radial velocities, as well as astrophysical information derived from the on-board multi-colour photometry.  This will allow the first three-dimensional map of our galaxy, and enable studies of its composition, formation and evolution.

During its lifetime, \gaia will be performing a full sky survey with an average cadence of 30 d. Around 80\% of observing periods will have at least two observations separated by 106 min.This turns \gaia into a tool for detecting a large number Galactic and extragalactic transient events. The focus of our study are low redshift supernovae (SNe) and tidal disruption events (TDEs). 

SNe are bright stellar explosions and according to the production channel, they may be classified as thermonuclear SNe type Ia, and core-collapse SNe (CCSNe). In the SN type Ia scenario, the explosion is triggered in a binary system, where a white dwarf (WD) is believed to have reached approximately the Chandrasekhar mass of 1.4 M${\odot}$. The WD could have accreted material from a non-degenerate companion or by the collision with another degenerate component. In the case of CCSNe, the explosion is caused by a gravitational collapse of a young and massive star, M$_{\rm init}\geq 8 $M${\odot}$ \citep{Smartt2009}. From the observational point of view, SNe have been historically classified into two main groups according to the presence or absence of hydrogen lines in their spectra \citep{Filippenko1997}. SNe type I, contain no hydrogen in their spectra. They can be further split into three sub-classes, those containing Si$_\mathrm{II}$ (type Ia), He $_\mathrm{I}$ (type Ib) or neither of them (type Ic). The hydrogen rich SN type II can also be further classified according to the width of the lines and light curve characteristics. With aim of simplification, we will consider all SN II as a single group and  make no distinction between types Ib and Ic, calling them just SN Ibc.

TDEs \citep{Rees1988} are flares originated from the total or partial disruption of a star orbiting at a close distance to the super massive black hole (SMBH) in the centre of its host galaxy. The circularization and accretion of the disrupted star creates a bright transient, often identified by its optical, UV and X-ray emission. The systematic study of TDE may probe the mass and spin distributions for the black hole population in non-active galaxies. Spectral and photometric evolution of the resulting flare provides valuable information on the debris circularization time-scale and the characteristics of the accretion processes in SMBHs.

The first TDE candidates found in an optical survey were discovered by \cite{vanVelzen2011}. Since then, the number of TDE discovered in optical surveys has increased rapidly \citep{ Gezari2012,Chornock2014,Arcavi2014,Holoien2014}. Our study, shows that \gaia could detect a significant number of TDE, substantially increasing the existing sample of this transient class. \gaia's excellent spatial resolution is key to distinguishing SNe from TDE, as the position of the transient with respect to the host galaxy nucleus may be used as a discriminant. This provides a model-independent way to select TDE. Current surveys mainly rely on light curve properties (e.g. shape and colours) and the spectra to select the candidates. However, this approach may bias the search outcome towards objects with similar properties. Some ground-based surveys that are sensitive to transients in the nuclei of galaxies, e.g. OGLE-IV survey for transients \citep{Wyrzykowski2014} have provided a number of atypical nuclear transients of yet not well understood origin.

The \gaia Photometric Science Alerts group \citep{Wyrzykowski2012} is the \gaia Data Processing and Analysis Consortium (DPAC) unit responsible for releasing alerts on \gaia transient candidates from a daily flow of initially reduced data. The Photometric Science Alerts pipeline builds on the Initial Data Treatment pipeline (developed by the DPAC team in Barcelona, with contributions from the teams in Leiden, Edinburgh and Lund) at the Data Processing Centre ESAC (DPCE) and uses the data ingestion facilities at the Data Processing Centre (Institute of Astronomy) Cambridge (DPCI). The alert flow is public (see \href{http://gaia.ac.uk/selected-gaia-science-alerts}{http://gaia.ac.uk/selected-gaia-science-alerts}) and it publishes a set of potential transient candidates, along with their characterisation using the low-resolution spectra provided by \gaia. The GS-TEC module \citep{Blagorodnova2014} is used to provide the classification type, redshift, and epoch of explosion for the most common SNe types. The transient candidates published by \gaia Photometric Science Alerts will enable the estimation of rates for several Galactic and extragalactic transient events. In order to provide reliable estimates of these rates, the biases in the \gaia detection process must be fully understood. 

Ground based transient surveys are generally facing two major issues: contrast against bright background and obscuration by host galaxy dust. The combined effect of seeing and imperfect image subtraction techniques makes the identification of events happening on top of bright galaxies challenging. The lack of atmospheric effects and the high spatial resolution of \gaia, comparable to the \textit{Hubble Space Telescope}, enables us to mitigate the first problem, as it allows transients to be resolved at closer angular separations to their host galaxy centres. In this work, we intend to quantify this effect in the circum-nuclear regime, where the numbers of detected SNe is lower \citep{Shaw1979}.  

The detection of SNe with high amounts of dust along the line of sight, such as SNe in interacting, or highly inclined spiral galaxies \citep{Cappellaro1997} will still remain a challenge for \gaia, which operates in the optical wavelengths. Several theoretical models  \citep{Hatano1998, RielloPatat2005} have been proposed to model the expected extinction in spiral galaxies as a function of inclination angle and the projected galactocentric distance. Moreover, the fraction of obscured star formation is known to increase as a function of redshift, e.g. \cite{Magnelli2014}. Recent studies using adaptive optics assisted near-IR observations have been able to detect and study SNe also within the nuclear regions of nearby luminous infrared galaxies (LIRGs), e.g. \cite{Kankare2012}. The introduction of corrections for missing SNe due to dust obscuration in the SN host galaxies of measured CCSN rates as a function of redshift, has improved the agreement with the rates expected from the cosmic star formation history \citep[see][]{ Melinder2012, Mattila2012, Dahlen2012}. In this work, we show that \gaia's superb high resolution imagery will enable it to detect a significant number of CCSNe. This sample will also allow new constraints to be set on the correction for the SNe missed due to dust obscuration, independently of the background light contamination bias.

This work is an attempt to understand the strengths and limitations of \gaia in relation to the detection of circumnuclear transients. Understanding the detection process and the impact of the selection procedure within \gaia Science Alerts is essential to provide reliable rates for the discovered transients. Understanding the parameter space for the detected candidates, will help to minimize the selection bias. Therefore, we aim to quantify the detection efficiency of transients located at few arcsec to their host galaxy nucleus. Our study particularly focuses on SNe and TDEs. Both types of objects are generally associated with a  host galaxy, either as an explosion of one of the stars in the galaxy, or its disruption by the SMBH existing in the host galaxy centre. The proximity of the host galaxy core is an observational effect that must be taken into account when computing the detection efficiency and predicted detection rates for transient surveys. Previous works have omitted this part of the analysis when providing estimates for the number of \gaia SNe \citep{BelokurovEvans2003, Altavilla2012}. This work, for the first time, quantifies the detection of the host galaxy and the contamination effect of the host on the detection of \gaia circumnuclear transients. Using a simulation of the on-board detection process and Monte Carlo (MC) simulations for generating SNe and TDE from a mock galaxy catalogue, we provide an estimate of \gaia 's detection efficiency and the predicted numbers of detected events for different candidate selection criteria.

The outline of the paper is as follows, Section \ref{sec:gaia_data} presents an overview of the \gaia data. Section \ref{sec:sims} explains the simulation techniques used in this paper. Section \ref{sec:results} presents the detection efficiency and rates results. Section \ref{sec:discussion} offers a discussion on the results with mission \gaia data and finally Section \ref{sec:conclusions} summarizes our conclusions.

\section{Description of the \gaia\ data} \label{sec:gaia_data}
 \gaia uses 106 charge-coupled devices, CCDs to record the data on an average number of $50\times 10^6$ transits a day. In order to reduce the data rate to a manageable level, around 50-130 Gigabytes per day \citep{Siddiqui2014}, only regions around detected sources are read out from the CCD and transmitted back to Earth.  The detection of sources must necessarily happen on-board, this is done using the Star Mapper (SM) CCDs and the sources are confirmed in the first column of Astrometric Field (AF) CCDs.  A full representation of the focal plane as well as the explanation of the mission pre-launch performance may be found in \cite{DeBruijne2012}.

In the standard \gaia terminology the area which is kept around the source, and eventually transmitted back to Earth, is referred to as a window. A window is composed of a set of CCD pixels, which may be binned together to form a macro-pixel. The size of the window, the number of pixels and binning depend on both the CCD and the magnitude of the object.

\gaia pixels are not square, but rectangular, the Along-scan (AL) direction is narrower, 0.059 arcsec, while the Across-scan (AC) direction is wider, 0.179 arcsec. In other words, the pixels are approximately three times larger in the AC than in the AL direction. This means that the characteristics of the source detection, in particular for asymmetric arrangements of emission, will depend on the orientation at which \gaia scans the source. 

\begin{figure}
\centering
\subfigure{\includegraphics[width=0.5\textwidth]{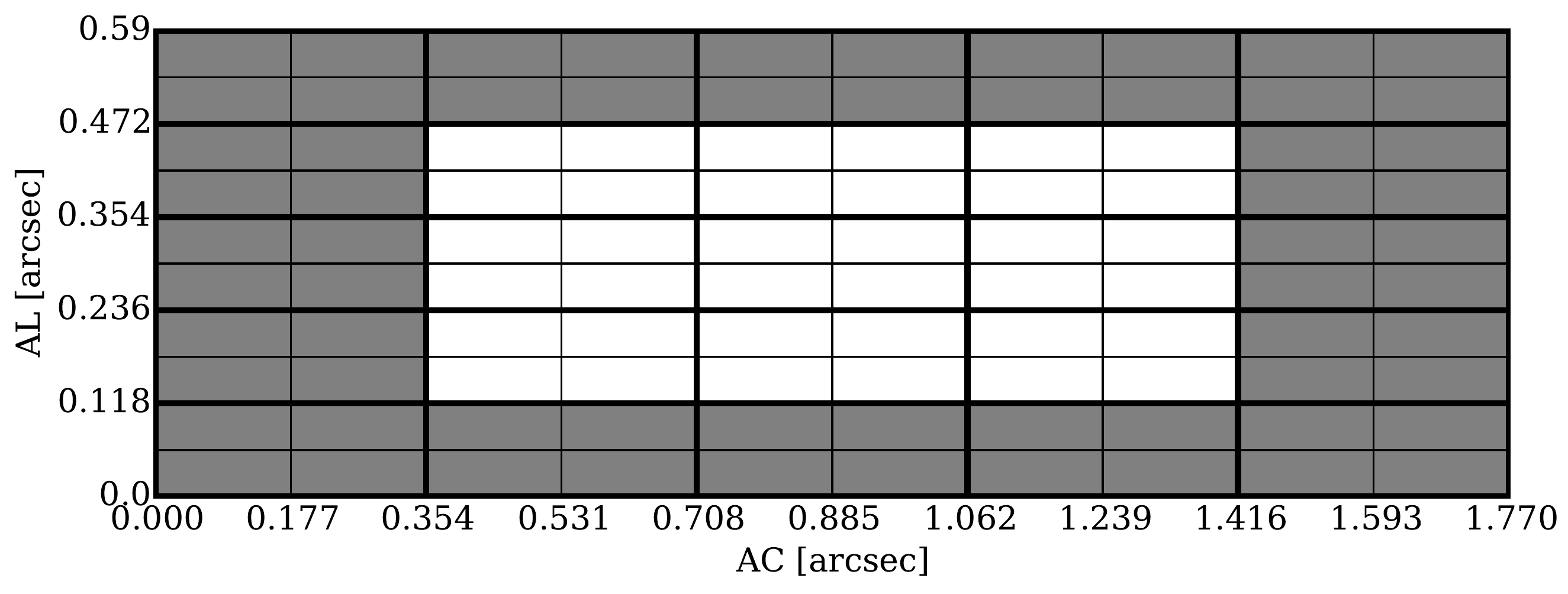}}
\caption{
SM window. The central white area of 3$\times$3 macro-pixels is used to estimate the flux of the source and provide an initial classification. The grey 5$\times$5 macro-pixels ring is used to estimate the background around the source. Thicker lines correspond to macro-pixels and the thinner lines to the pixels in the SM window. The $Y$ axial corresponds to the satellite scanning direction.}
\label{fig:sm_samples_fig}
\end{figure}

The detection process takes place in the 0.59 $\times$ 1.77 arcsec SM working window represented in Figure~\ref{fig:sm_samples_fig}, where pixels are hardware-binned $2\times2$. The central region of 9 macro-pixels is used to estimate the flux of the object, while the external ring of 5 macro-pixels is used for background subtraction. The SM window sent to Earth around the position of the detection is larger though, covering 4.72 $\times$ 2.12 arcsec. A detailed description of the detection process and the recognition strategy used in the on-board Video Processing Algorithm (VPA) may be found in de  \cite{deSouza2014} and \cite{DeBruijne2015}.

The spin rate of {\it Gaia} is 60 ${\rm arcsec \ s}^{-1}$ and the axial maintains an angle of  $45^{\circ}$ to the Sun, while slowly precessing around the solar direction, completing a full revolution every 63 d.  After its 5-year mission each object will have been observed between 50 and 250 times depending on the position of the source on the sky, with the ecliptic latitude of the source  being the most important factor in determining the coverage.

\section{Simulations} \label{sec:sims}

We took advantage of the functionalities provided by the \textsc{gibis} simulator (\gaia Instrument and Basic Image Simulator; see Appendix \ref{sec:gibis} for a description) to assess the characteristics of the on-board detection.  We used a mock galaxy catalogue to reproduce the host galaxy characteristics and empiric light curves to mimic the observational signature of SNe and TDE in the optical wavelengths. Details on our models and assumptions are provided in Appendices \ref{sec:galmod} and \ref{sec:tranmod}.

In this section we describe the simulations and analysis techniques that we use to obtain the transient detection efficiency for \gaia. We first obtain the on-board characteristics of the detection process for galaxies and for transients close to nucleus. Later we use these results in MC simulations to determine the detection efficiency for the whole survey.

\subsection{Galaxy detection simulation} \label{sec:galaxy_simulation}

The detection efficiency of transients is constrained by the detectability of the host galaxy. Moreover, to accurately estimate the distance of a transient to the nucleus of its host galaxy, the astrometry for both components needs to be delivered by \gaia, and therefore the galaxy itself has to have the appropriate characteristics to be detected by the on-board detection algorithm. According to the galaxy detectability study in \cite{deSouza2014}, \gaia has a detection process optimised for point-like sources, leading to a selective galaxy  detection, dominated by compact galaxy bulges or small elliptical galaxies. Therefore, our analysis focused on the bulge component only.

We chose to parametrize the host galaxy bulges using two main observables: bulge apparent magnitude, $m_{\rm G}$, expressed in \gaia $G$-magnitudes, and the galaxy bulge angular size, characterized by the bulge effective radius, $r_\mathrm{e}$. See \ref{sec:galaxy_simulation_conf} for simulation configuration details.

\subsection{Transients detection simulation} \label{sec:transient_simulation}

Using \textsc{gibis} we simulated transients next to galaxy cores. The transients had a wide range of apparent magnitudes $m_\mathrm{T}$ and they were placed at increasing angular separation, $\theta$, from the centre of their host galaxy. 

These simulations provide the \gaia detection parameters for a set of possible transient configurations. The initial result is the probability of detecting the transient on board given the parameters defining the host galaxies and the transient.

\begin{equation}
P( \texttt{det} | m_{\rm G}, r_{e}, m_{T}, \theta)
\label{eq:det}
\end{equation}

In case of a successful detection, the second result is the probability that the transient has generated an additional new detection, meaning that it has been resolved into a separate window.
\begin{equation}
P( \texttt{new} | m_{\rm G}, r_{e},  m_{T}, \theta,  \texttt{det} )
\label{eq:new}
\end{equation}

Finally, the third result is the transient magnitude, $m_{\rm VPA}$ as predicted by the on-board algorithm, the VPA, provided that it was detected.
\begin{equation}
P(m_\text{VPA} | m_{\rm G}, r_{e}, m_{T}, \theta,  \texttt{new},\texttt{det}  )
\label{eq:detmag}
\end{equation}

\begin{figure}
\hspace{-0.1cm}
\includegraphics[width=1.0\columnwidth]{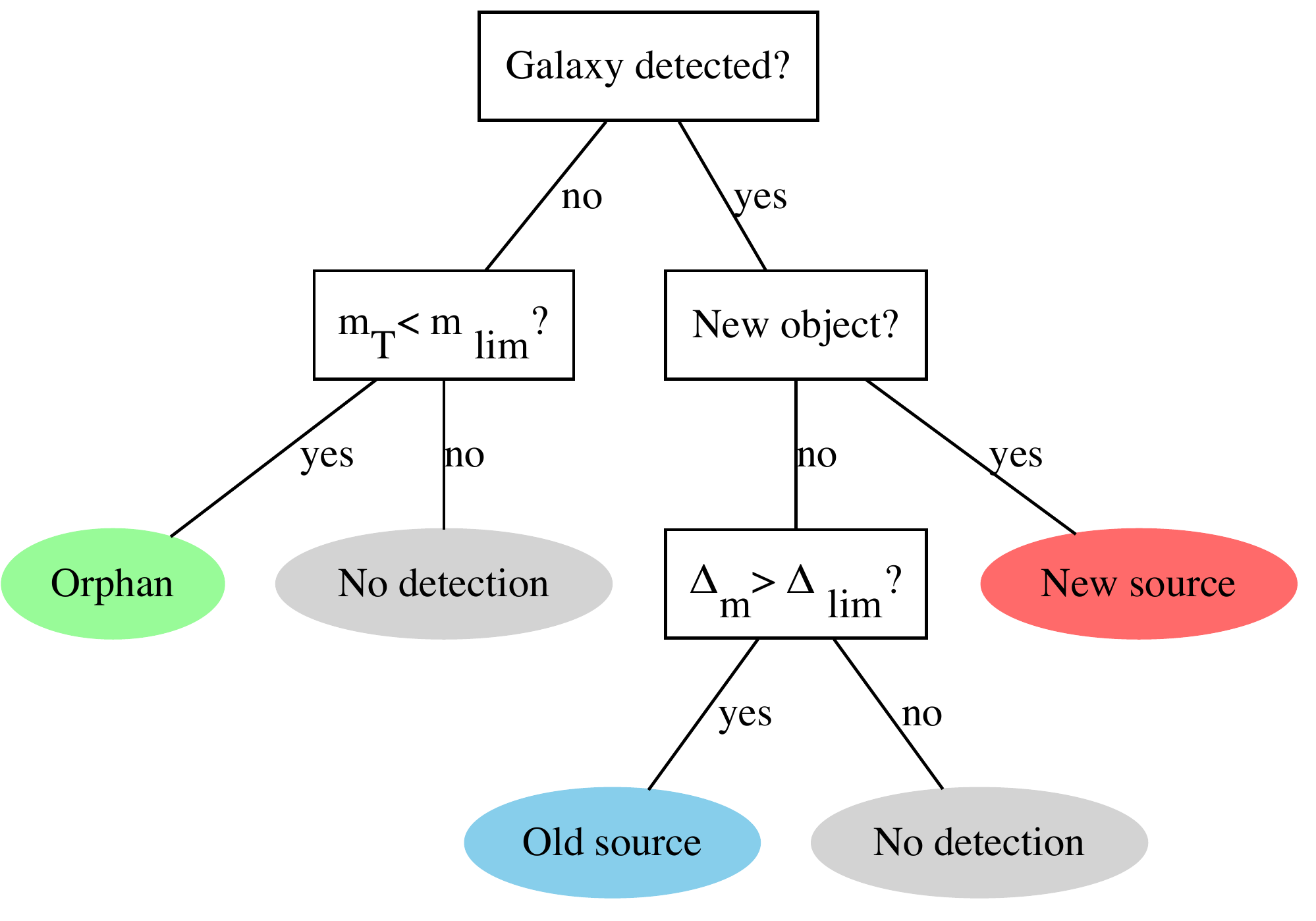} 
\caption{Transient alerts detection process.  For non-detected galaxies, new sources will be created whenever the transient magnitude, $m_\mathrm{T}$, is brighter than the alerting limiting magnitude $m_{\rm lim}$. For detected galaxies, source detection is based on the spatial resolution of the object. Resolved objects will be treated in the same way as new sources, and old objects will only produce alerts if the magnitude of the host galaxy increases more than $\Delta {\rm lim}$, due to the transient light contribution.}
\label{fig:det_diagram}
\end{figure}

\subsection{Candidate selection} \label{sec:candidate_selection}

The \gaia Science Alerts pipeline runs the candidate detection process for each 24 h of initially reduced \gaia observations. The outcome is a list of an initial selection of transient candidates. The assumed on-ground candidate selection process is shown in Fig. \ref{fig:det_diagram}. This process has two main parameters: limiting magnitude for transient alerts, $m_{\rm lim}$, and the minimum change in magnitude, $\Delta $m, for old existing sources.

The selection of $m_{\rm lim}$ is fixed at $G=19$ in order to minimize false positives. This means that transients only with detections above 19th magnitude will be recognized and published as such. In this context, $m_{\rm lim}$ does not refer to the limiting magnitude of the \gaia mission, which is located between 20 and 21st mag, but rather to the limiting magnitude for selecting and publishing alerts candidates. Further into the mission, this limit could be relaxed, allowing the publication of fainter detections.

The selection of $\Delta$m was also chosen to reduce the number of false positives. The \gaia photometric error per transit is expected to be lower than 0.01 mag in $G$ band for single 20 mag stars \citep{DeBruijne2012}. However, the scanning-angle dependency for close unresolved stars and highly elliptical galaxies may introduce additional variability into the data. The selection of $\Delta$m has also an implication on the selection of quasars. For a given quasar, we expect that it has a probability of 1\% to vary by $\Delta\mathrm{m} > 0.3$ within a period of 30 d and by $\Delta\mathrm{m} > 0.5$ within a period of 100 d \citep{MacLeod2012}.

The strategy used to select transient candidates is temporary and may change in the future. As of mid 2015, each source needs at least three historical data points (detections or non-detections), in order to provide a baseline to monitor its magnitude. 

For transients in non-detected host galaxies, they will only be selected if the transient magnitude is brighter than $m_{\rm lim}$ and there are at least two data points above this limit. The alert candidates will be new sources on their own, as no previous records on their host will be available in \gaia catalogue.  We will call them \textit{Orphan}.

On the contrary, transients in galaxies detected by \gaia, are either resolved into a new independent detection or are detected along with the galaxy core. For new sources, the same criteria as before will be applied, requiring an apparent magnitude above the limiting threshold and at least two data points. Given that the nuclei of their host galaxies have been identified and resolved from the SN, we call them \textit{New Source} detections. Otherwise, if the transient is too close to the host or too faint to be resolved, then what will trigger the alert is an overall change in brightness of an already catalogued source, if this change is larger than the selected detection limit, $\Delta\mathrm{m}$. For this case, as well, two data points are required for a robust detection. We will refer to these cases as \textit{Old Source} detections. Although we call them \textit{unresolved}, these kind of transients, if caused by off-nucleus SNe, will generally display a detectable offset shift in the position of the light centroid, which can be determined with an accuracy of 100 milliarcsec or better in the daily reduced \gaia data provided by One Day Astrometric Solution.

\subsection{Detection efficiency calculation} \label{sec:method}

\subsubsection{TDE} \label{sec:method_tde}

Given the generated galaxy catalogue, described in Appendix \ref{sec:galmod} B1, and the TDE light curves, described in Appendix \ref{sec:tranmod}, the simulation of the detection process was performed in the following way. For each galaxy in the catalogue, a TDE disruption was simulated at random sky coordinates at a random day after the start of the mission. The light curve used to model the optical emission of the TDE was selected from two different light curves groups: Pan-STARRS TDE candidate events (PS1) and TDE light curve models from \cite{LodatoRossi2011} (LR11). These are further described in the Appendix \ref{sec:tranmod}. The Milky Way extinction along the line of sight has been provided by the Galactic extinction map from \cite{Schlegel1998} with the correction provided by \cite{Bonifacio2000} and $R_\mathrm{V}=3.1$. Reddening due to extinction in the TDE host galaxy is not included, as the TDE rates we use are based on detected events, which do not include reddening corrections.

Each TDE light curve was sampled according to the sky position and the Nominal Scanning Law (NSL) of \gaia. Then, each sampled data point was assigned a probability for having been detected on-board using \textsc{gibis} results described in Section \ref{sec:transient_simulation}. We interpolated the discrete outcome of the set of simulations with parameters of $m_{\rm G}$, $r_\mathrm{e}$ and $m_\mathrm{T}$ to provide a probability of detection and observed magnitude for each point in the light curve. Finally we applied the transient selection criteria (i.e. cuts in $m_{\rm lim}$ and $\Delta\mathrm{m}$) to compute how many of the TDE candidates would be detected by the pipeline. We run this MC simulation on a sample of $1.5 \times 10^6$ galaxies.

\begin{figure*}
\centering
\mbox{
\subfigure{\includegraphics[width=1.\columnwidth]{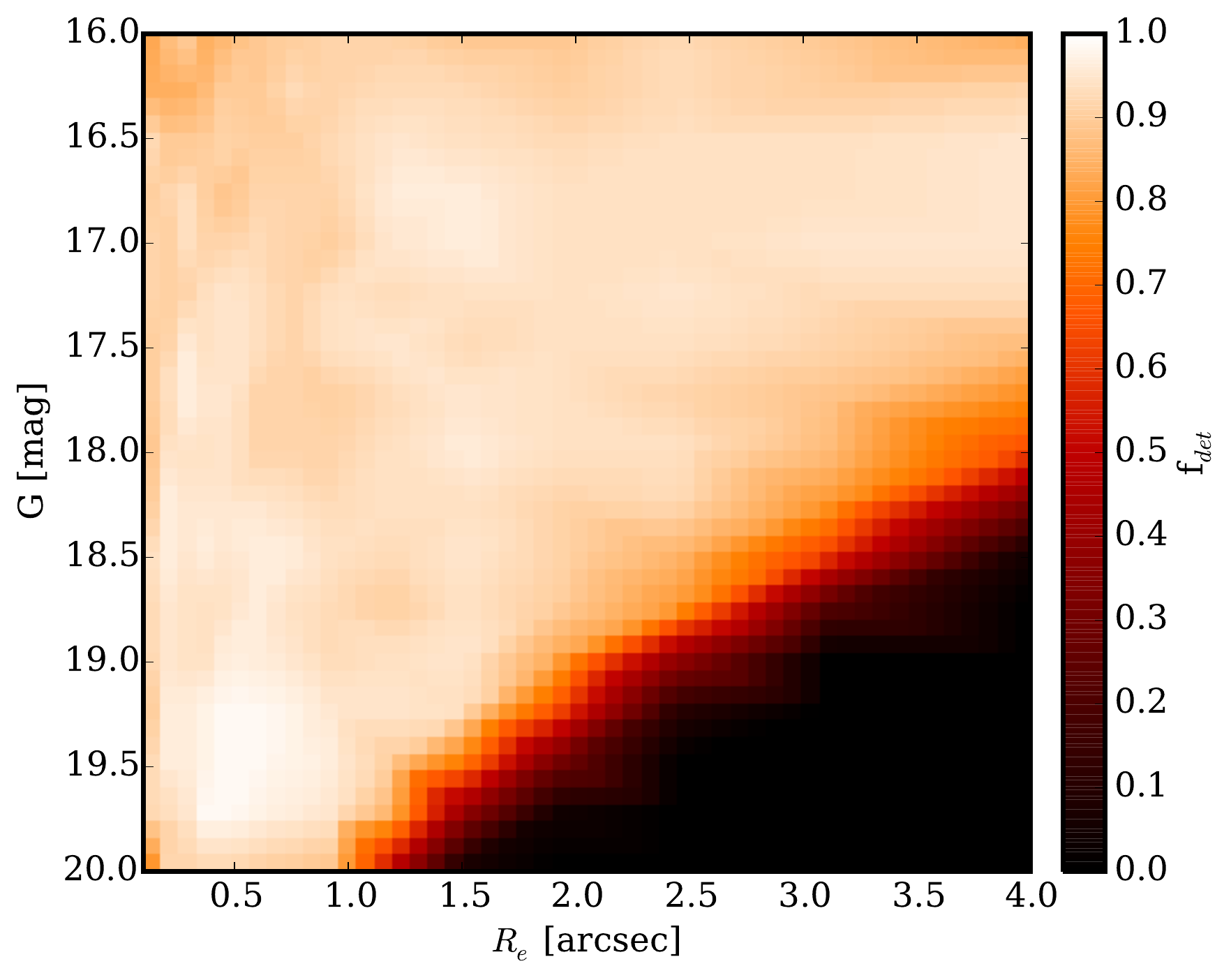} 
\quad
%\hspace{0.5cm}
\subfigure{\includegraphics[width=1.\columnwidth]{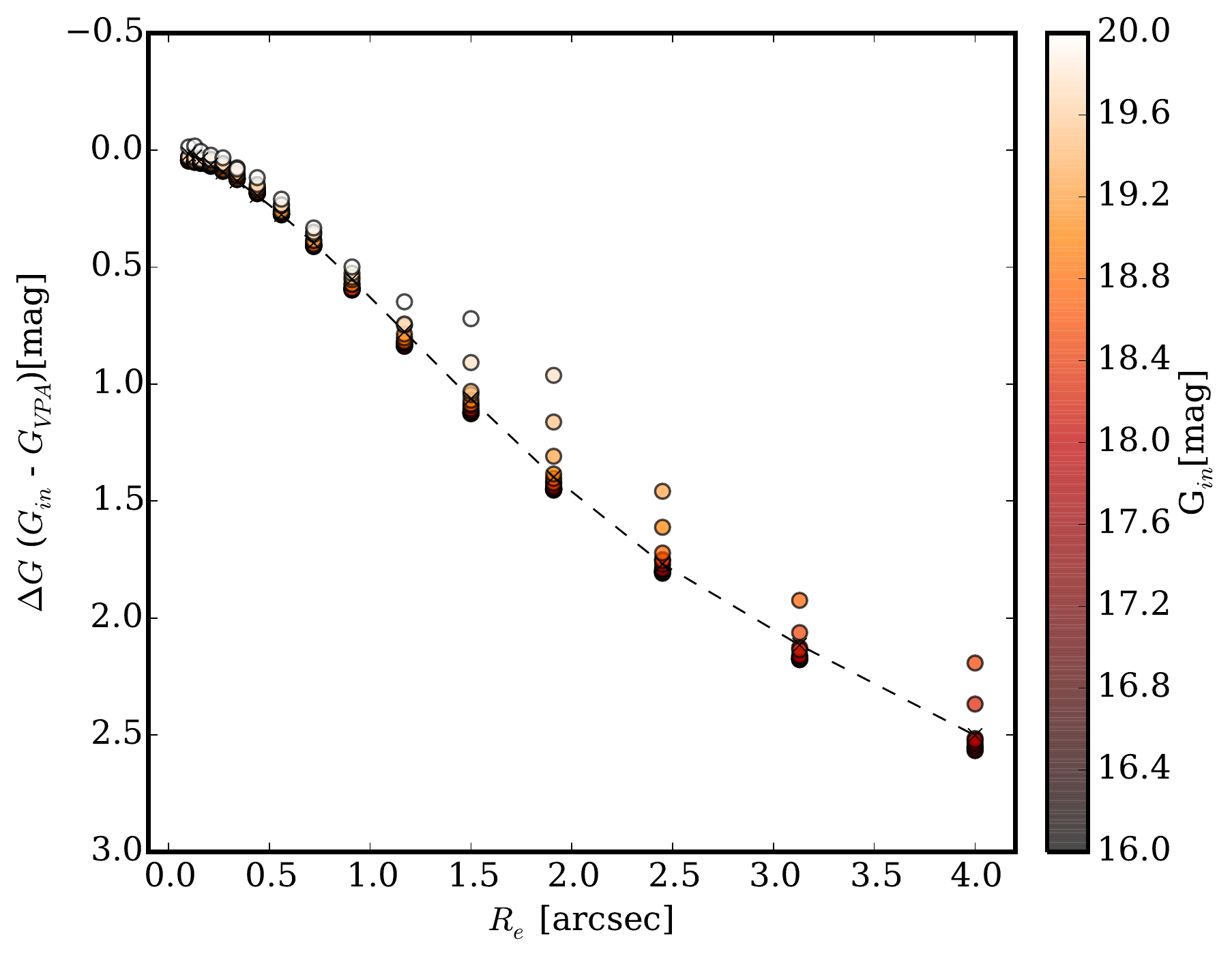}}} }
\caption{ Left:  \textsc{gibis} simulation results for detection of galaxies with $B/T$=1 and $b/a$=1. The colour code shows the detection probability for galaxies with sizes $r_\mathrm{e}$=(0.01--4.) arcsec and magnitudes $G$=(16--20.0). The detection probability is computed as the fraction of scans when the simulated objects were detected over the total number of scans. Objects falling in between CCD gaps are accounted as non-detections. Each scan was generated at different inclination following the Nominal Scanning Law. Right: difference between the input and simulated magnitude as function of the bulge effective radius $r_\mathrm{e}$. $G_{\rm in}$ represents the integrated (true) magnitude used as input in the galaxy bulge simulation, and $G_{\rm VPA}$ is the magnitude estimated by the on-board detection algorithm. The dimming effect mainly depends on the size of the galaxy, which is translated in flux loss outside of the detection window. The dependency on bulge magnitude (colour coded) is weak. Bulges with fainter magnitudes progressively disappear from the relation, as they stop being detectable with higher values of $\Delta G$. The black dashed line shows the best-fitting four degree polynomial $f(x) = 0.022-0.187x+0.392x^2+ 0.794x^3 -0.080x^4$.}
\label{fig:gal_rac_det}
\end{figure*}

After obtaining the detectability results from the simulations, we computed the total number of detected TDE in the following way. For a given redshift bin, the volume density of detected TDE is given by:

\begin{equation}
\rho(z)  = \sum^T  \int \tau \phi^*(T, M_r) \varphi(T, M_r, z) \alpha(M_r)\  \mathrm{d}M_r,
\label{eq:det_eff_z}
\end{equation}

where $\phi^*(T,M_r)$ is the galaxy luminosity function in r-band from \cite{Baldry2004} for galaxy type $T$ (red or blue) and $\tau$ is the average TDE rate per galaxy per year. Following the approach described in \citep{vanVelzenFarrar2014}, we assume two different TDE rates, according to the chosen modelling of the light curves. We use $\tau= 2.0\times 10^{-5}$ TDE yr$^{-1}$ galaxy$^{-1}$ for PS1 light curves and a rate of $\tau=1.7 \times 10^{-5}$ TDE yr$^{-1}$ galaxy$^{-1}$ for the LR11 model light curves. We define $\varphi(T, M, z)$ as the detection probability for TDE hosted by galaxies with the given absolute magnitude, type and redshift bin and $\alpha(M_r)$ is the probability that galaxies of given absolute magnitude could generate a TDE flare, instead of totally swallowing the star \citep{Kesden2012}. Assuming that the probability of TDE suppression follows an exponential law driven by the black hole mass:

\begin{equation}
\widetilde{\alpha}(M_{\rm BH}) = \text{exp} (-M_{\rm BH}/3 \times 10^7 M_{\sun} )^{0.9}
\label{eq:supress_rate}
\end{equation}
we can transform it to a suppression per galaxy luminosity using bulge mass to galaxy luminosity transformation, $M_r=f(M_{\rm BH})$. Then, we can transform $\widetilde{\alpha}(M_{\rm BH}) =  \alpha(f(M_{\rm BH})) = \alpha(M_r)$.

Finally, the expected number of detected TDE is computed by integrating the density of detectable TDE for each redshift bin from Equation \ref{eq:det_eff_z} over the differential volume at each redshift:
\begin{equation}
N_\text{\rm TDE} \text{ yr}^{-1}=   \int    V_\mathrm{C}(z)  \ \rho(z) \ \mathrm{d}z
\label{eq:ntde}
\end{equation}
where $V_\mathrm{C}(z)$ is the comoving volume for redshift $z$.

\subsubsection{Supernovae}\label{sec:method_sn}

The simulation for the detection process for SNe is carried out in a similar way to that of TDE. See Appendix \ref{sec:tranmod} for assumptions on SN rates and light curves. Here we also generate random sky positions for the hosts and random times for the epoch of explosion. As mentioned in Appendix \ref{sec:tranmod}, the absolute magnitude distribution and the SN light curves are obtained from \cite{Li2011}, and do not include a correction for host galaxy extinction. We therefore also avoid applying any host galaxy extinction for the simulated objects, as this would create SNe that have been dimmed twice: for each SN we only apply the Milky Way extinction.

We simulate thermonuclear SNe in both early-type and late-type galaxies, as they are found among both old and young stellar populations \citep{Maoz2011}. The distribution of distances follows the overall stellar mass in the galaxy (\cite{ForsterSchawinski2008}, \cite{JamesAnderson2006}), and therefore their distances are uniformly drawn following the combined light profile of bulge and disc. 

CCSNe on the contrary are only simulated in late type galaxies, as they originate from the young, massive star population. The deaths of short-lived stars closely follow the star formation regions \citep{JamesAnderson2006}, which are located in the galaxy discs. Consequently, the radial distance distribution of CCSNe is simulated following the disc light only. With the aim of simplification, we do not include starburst galaxies or LIRGs in our simulations, which show evidence of more centrally located population of CCSNe \citep{Herrero-Illana2012}.

The main parameters used for the simulation of the SNe are the angular separation parameter, $\theta$, the host galaxy magnitude, $m_{\rm G}$, and angular size, $r_\mathrm{e}$, and the magnitude for each sampled point in the SN light curve, $m_{\rm SN}$. For a given combinations of these parameters, obtained from the MC simulations, we check the on-board detectability of each point and we select the transient candidates using the process explained in Section \ref{sec:candidate_selection}.  

Computing the number of detected SNe of type X (SN Ia, SN Ibc or SN II) follows the same procedure as for TDE. First we derive the volume density of SNe of type $X$ that are detected for each redshift bin:

\begin{equation}
\rho (X, z)  =\sum^T  \int^{M_r} \phi^*(T, M_r)\text{SNuB}(T,X,M_r)  \varphi(T, M_r, z) \ \mathrm{d}M_r
\label{eq:det_eff_z_sn_type}
\end{equation}
where $\phi^*(T,M_r)$ is the galaxy luminosity function in r-band, SNuB$(T,X,M_r)$ is the rate of SNe type $X$ per galaxy type $T$ and absolute magnitude $M$ and $\varphi(T, M_r, z)$ is the detection probability for SNe hosted by galaxies with the given absolute magnitude, type and redshift. The expected number of SNe is obtained by multiplying the volume density of SNe yr$^{-1}$ by the volume in each redshift bin $dz$.

\begin{equation}
N_{X}\text{ yr}^{-1}=      \int^z   V_C(z) \rho(X,z) \ \mathrm{d}z
\label{eq:per_galaxy}
\end{equation}

\begin{figure*}
\centering
\subfigure{\includegraphics[width=1.4\columnwidth]{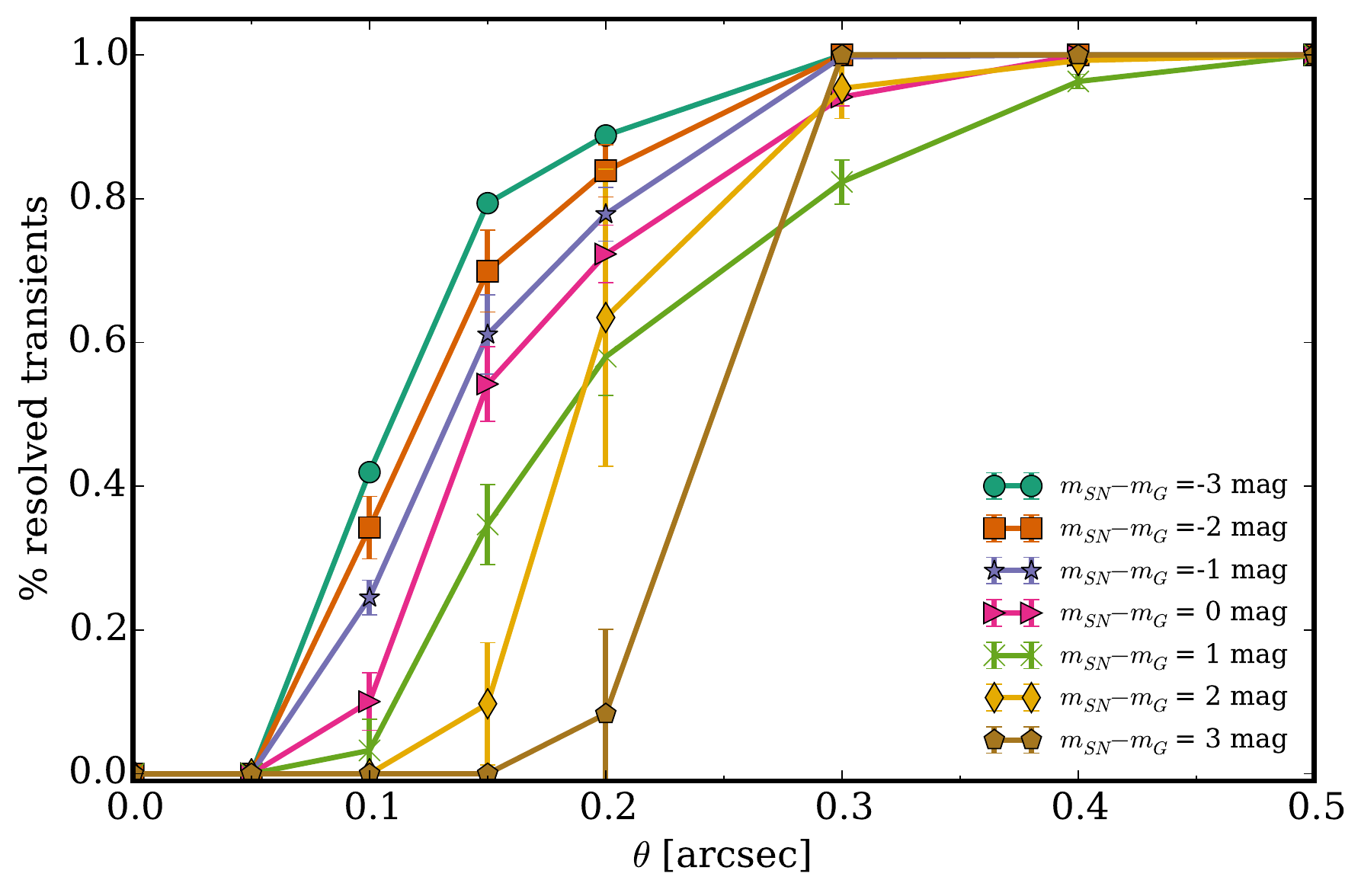} }
\mbox{
\subfigure{\includegraphics[width=1.05\columnwidth]{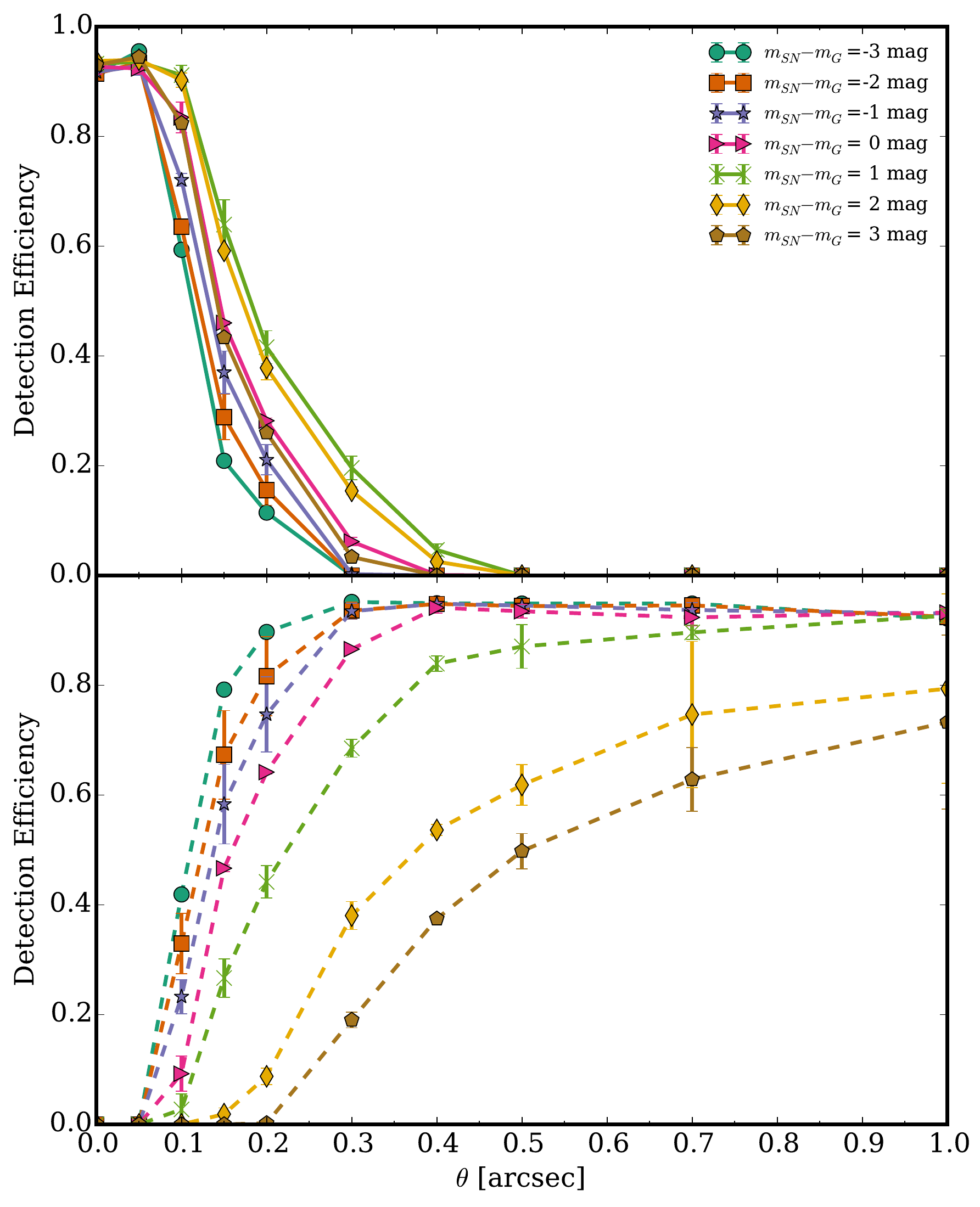} 
\quad
\hspace{-0.5cm}
\subfigure{\includegraphics[width=1.05\columnwidth]{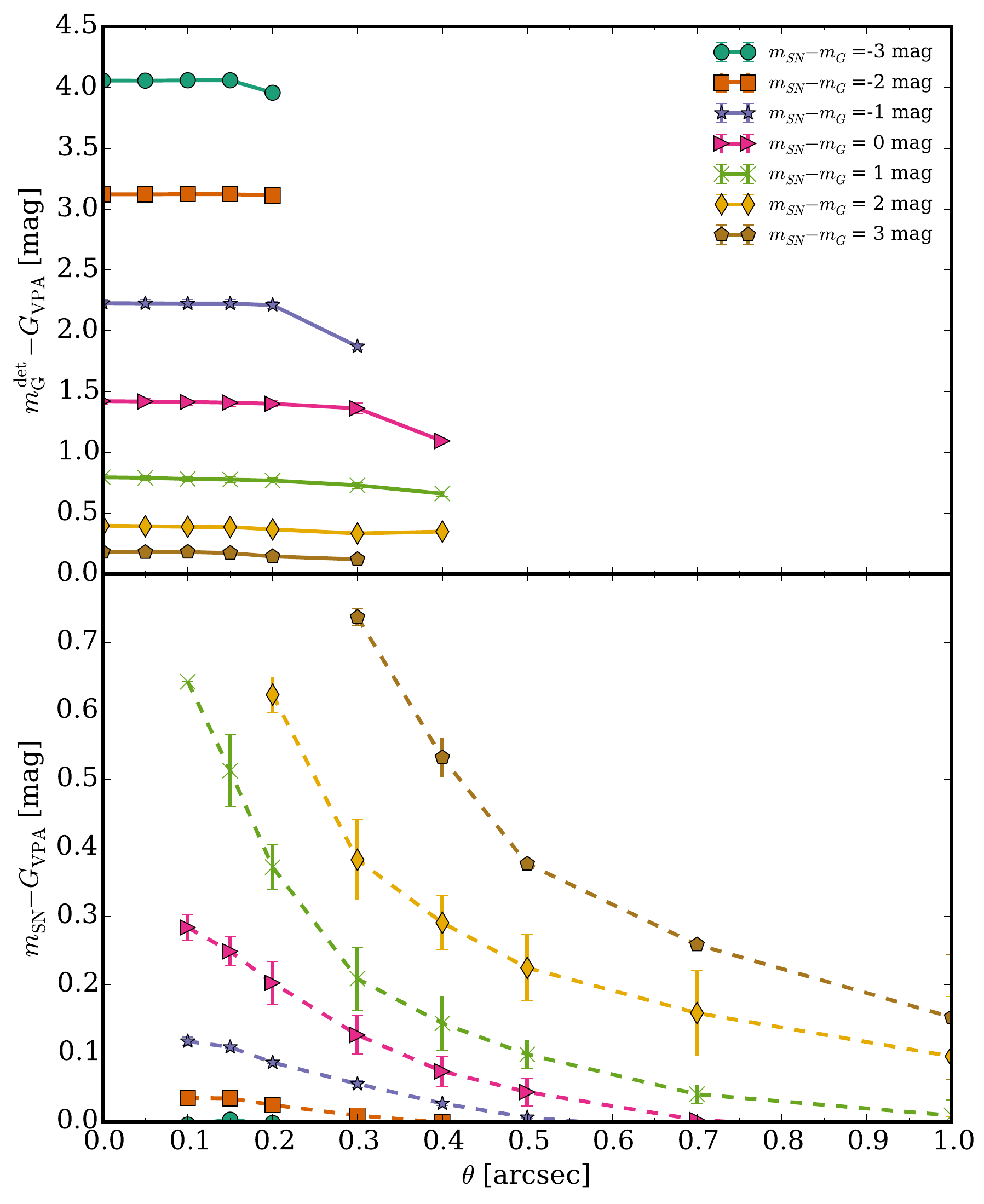}} } }

\caption{Top: \textsc{GIBIS} simulation results for transients in galaxies with $B/T$=1, $r_\mathrm{e}$ =1 arcsec, describing elliptical galaxies or compact bulges. Top: From the sample of detected SNe, the fraction of resolved ones as function  of their angular separation, $\theta$, and difference in magnitude  $m_{\rm SN} - m_{\rm G}$ (line colour). SNe which are 2--3 mag fainter than the host, have a low probability to be detected on board (see bottom left panel). The detection only happens when the scan orientation is favourable to resolve the SN from the host. Bottom Left: on-board detection efficiency for transients as function of angular separation and difference in magnitude $m_{\rm SN} - m_{\rm G}$ (line colour).  Dashed lines represent the detection probability for resolved objects and the solid lines for unresolved objects. The curves are averaged values for bulges of magnitudes 16 to 20. The error bars are determined by the scatter on the average computed for each bin. The results are only shown for bulges with an effective radius ($r_\mathrm{e}$) of 1 arcsec for clarity. The detection probability is computed as the number of detections over the total number of scans (or potential detections).  Due to different scanning angles, some of the simulated object may fall into gaps between the CCDs. Therefore, the detection efficiency per simulated object is lower than 100\%.
 Bottom Right: difference in magnitude between the transient input magnitude, $m_{\rm SN}$, and the on-board detected magnitude, $G_{\rm VPA}$, for bulge-transient pairs as function of angular separation. Same magnitude difference bins are used as in the left subfigure. Dashed lines represent the magnitude change for resolved objects and the solid lines for unresolved objects.}
\label{fig:det_eff_sn}
\end{figure*}

\section{Results} \label{sec:results}

\subsection{On board galaxy detection} \label{sec:on_board_galaxy}

The probability for the on-board detection of the host galaxy, as determined by the \textsc{gibis} simulations described in Appendix \ref{sec:galaxy_simulation_conf},  is given by the fraction of scans in which the galaxy was detected over the total number of scans.
The detection fraction, $f_{\rm det}$, is shown in Fig. \ref{fig:gal_rac_det} as a function of two main parameters: bulge magnitude $m_{\rm G}$ and angular size, $r_\mathrm{e}$. The simulation was run in a representative region in terms of transit coverage as described in Section \ref{sec:galaxy_simulation}. Tests with the same configuration but different coordinates did not change significantly the detection probability results. Therefore, we conclude that the chosen position might be used as a representative case for all sky detection. 

Additional tests were done to assess the impact of changing the colour ($V-I$) and the axial ratio ($b/a$) for the simulated objects.  Variations of $V-I$ in the range 0--1.5 lead to no change of the detection efficiency. Results for tests on galaxy detection varying the ellipticity or axial ratio $b/a$ have shown that more elongated bulges are more likely to be detected at fainter magnitudes than spherical ones, as their light profile is more concentrated. This effect would increase slightly (around 30\%) the detection of galaxies in the limiting positions of the ($m_{\rm G}$, $r_\mathrm{e}$) parameter space. However, this effect is well below statistical uncertainty when we compare the total number of detected galaxies. With the aim of simplification, we considered that if the majority of bulges have lower ellipticities around $b/a\simeq$0.7 \citep{Maller2009}, and there is no significant difference between using the \textsc{gibis} detection function for $b/a$=1 from $b/a$=0.7, we could use $b/a$=1 for the general bulge population.

In order to obtain the total estimated number of detected galaxies, for each redshift and magnitude bin, we computed the galaxy detection efficiency using the \textsc{gibis} results obtained above and we integrated over the red and blue galaxy luminosity functions from \cite{Baldry2004} and redshift range (0--0.12), as galaxies further than 600 Mpc are unlikely to be detected \citep{deSouza2014}. This calculation shows that around $1.5 \times 10^6$ red and around $8 \times 10^4$ blue galaxies will be detected by \gaia.  Therefore, a total number of $\sim 1.5 \times 10^6$ galaxies is expected, which is a factor 3 higher than the numbers predicted by \cite{deSouza2014}. 

The main reason for the discrepancy is our assumption that all bulges of red galaxies have a S\'ersic index of 4, which is not always true. Lower indices are more common for describing the shallower light profile for bulges of S0 type galaxies \citep{Graham2001}. Given that bulges with lower indices are less likely to be detected by the on-board algorithm, our estimation is an upper limit for the number of detected galaxies. However, according to the last update of the software for the \gaia VPA, the new on-board detection algorithm has been tuned to detect bulges with shallower light profiles \citep{DeBruijne2015}. The impact of this update is an increase in the galaxy detection efficiency for bulges with lower S\'ersic indices and therefore providing better agreement with our results.

The analysis of the detected magnitude as function of bulge angular size is shown in the right panel of Fig. \ref{fig:gal_rac_det}. We can see the difference between the input magnitude $G_{\rm in}$ and the magnitude estimated on-board, $G_{\rm VPA}$ starts has an almost linear trend with the bulge angular size. Given that the flux is estimated under the assumption of point-like objects, galaxies with extended radii will be detected as much fainter objects. The figure also shows that there is a magnitude cut-off close to $G_{\rm VPA}=20$, so that objects with an estimated magnitude fainter than this limit are no longer detected. As mentioned above, generally galaxy bulges and pseudobulges will have a shallower light profiles than De Vaucouleurs' profile, and therefore will appear much fainter than in our simulations. Therefore, the difference in magnitude displayed here is also an upper limit for bulges with S\'ersic indices $n<4$.

\subsection{On-board transient--galaxy identification} \label{sec:on_board_transient}

The uneven resolution of \gaia's rectangular pixels creates a scanning angle dependency when resolving two nearby objects (binary star or a galaxy core and a SN). Scanning along the separation axial will provide enough accuracy to differentiate the objects, which, if resolved on-board, will be tagged as two separate detections with an allocated window each. Contrarily, scanning across the separation axial will increase the chances of the two objects being blended, resulting in a single detection with enhanced flux. Whenever the galaxy bulge and the SN are separated by less than 1 SM pixel, there will be only one detection on-board and the magnitude will be determined from the combined emission from the galaxy and the SN. 

Following the configuration explained in Appendix \ref{sec:tran_simulation_conf}, we run the on-board detection for the bulge-transient systems using the \textsc{gibis} simulator. Each combination of the four main parameters $ \{m_{\rm G}, r_\mathrm{e}, m_{\rm T}, \theta \}$ (bulge magnitude, bulge size, transient magnitude and angular separation between the bulge and the transient) provided three main results, formalized in Equations \ref{eq:det}--\ref{eq:detmag}. We can interpret these \textsc{gibis} outputs as functions, which for a given set of parameters, return the detection probability, the probability to resolve the system and the on-board estimated magnitude for the transient.

The detectability results are summarized in Fig. \ref{fig:det_eff_sn}, corresponding to simulations of galaxies with $r_\mathrm{e}=1$ arcsec. The top panel shows the fraction of resolved objects over the total number of simulated objects as a function of their angular separation. Objects of similar magnitudes ($\pm$1 mag difference) are less likely to be resolved, as they display a more uniform light distribution. The largest angular separation for unresolved objects lies between 0.4 and 0.5 arcsec. We can conclude that all objects located beyond 0.5 arcsec from their hosts, if they are detected, are always going to be resolved objects. 

The probability of detection as function of the angular separation is showed in the lower left panel. The unresolved (top) and resolved (bottom) cases are shown per separate. We bin the simulation results according to the difference between the transient's magnitude and the bulge magnitude $m_{\rm SN} - m_{\rm G}$.  Brighter transients ($m_{\rm SN} - m_{\rm G} < 0$) or about the same magnitude than their host's bulge are more likely to be detected at closer angular separations. Fainter transients ($m_{\rm SN} - m_{\rm G} > 0$) require a minimum angular separation between 0.2 and 0.3 arcsec to start being detected as resolved objects, and separations of 0.5 - 0.7 arcsec to have at least 0.5 probability of being detected. Finally, the detection efficiency never reaches 1, due to the gaps between the CCDs. From the initial set of the simulated objects, not all of them will be observed for a particular scanning angle. 

The proximity of the bulge also has an effect on the transient's estimated magnitude, as shown in the lower right panel of Fig. \ref{fig:det_eff_sn}. Transients unresolved from their hosts (solid line) are detected as the combined flux from both sources. Given that galaxies have a flux loss due to the limitation of window size in \gaia (see Fig. {\ref{fig:gal_rac_det}), the contribution from the SN may substantially increase the detected magnitude of the galaxy. For example, for bulges with $r_\mathrm{e}=1$ arcsec, SNe with $m_{\rm SN}-m{\rm G}=1$ already may create an increment in the detected bulge magnitude of $\Delta\mathrm{m} = 0.8$.

For transients having the same magnitude than their hosts, the bulge light contribution makes them about 0.5 mag brighter. For resolved transients (bottom), the light contribution from the bulge exponentially decreases at larger angular separations. From these results we may conclude that transients located further than 1 arcsec, will barely suffer any contamination from their hosts. Transients at angular separations closer than 1 arcsec, will appear brighter than they are, as they will include additional light from their hosts.

\subsection{Survey detection efficiency} \label{sec:ground_detection}

\begin{figure}
\centering
\includegraphics[width=\columnwidth]{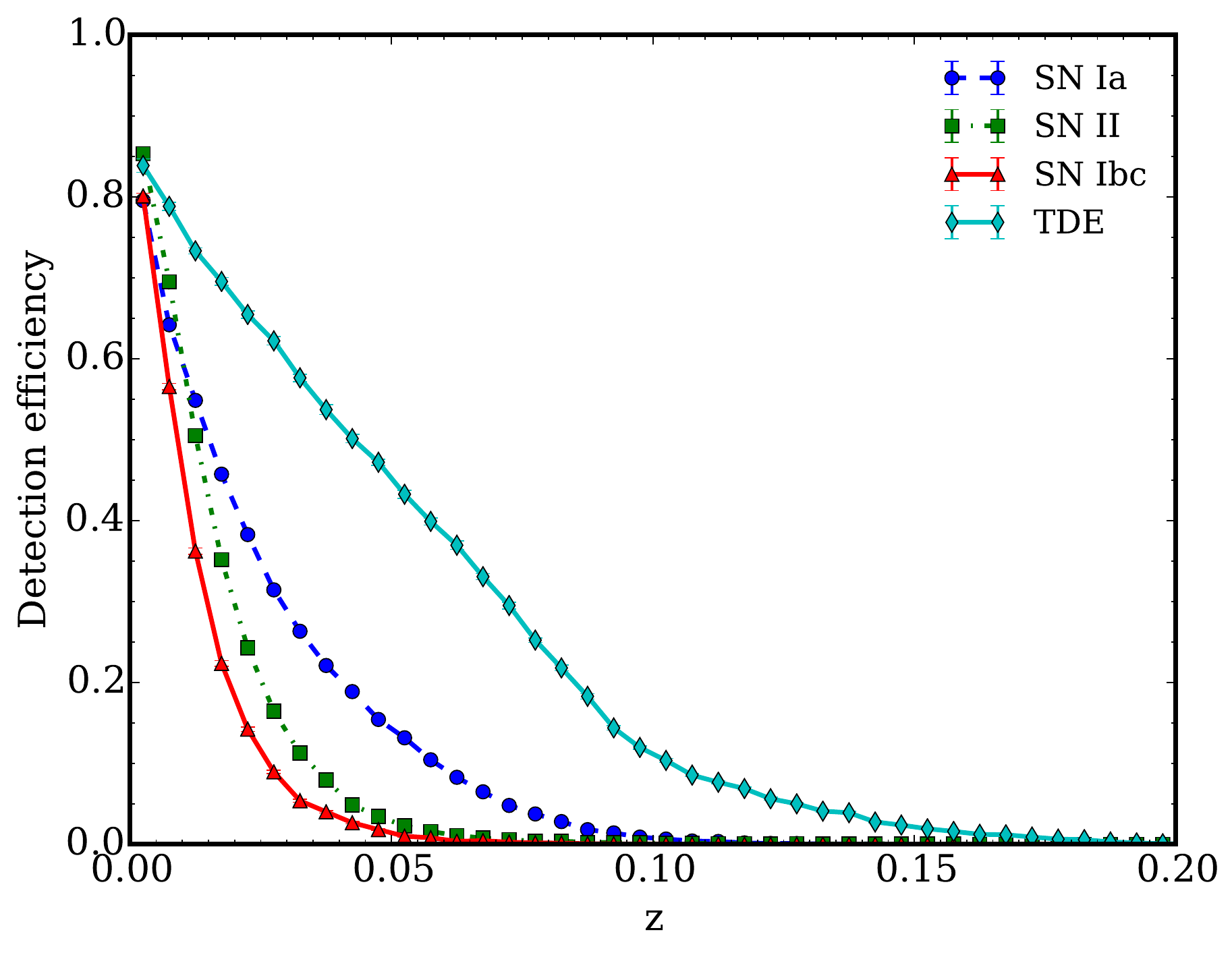} 
\caption{Detection efficiency for each redshift bin for $m_{\rm lim}=19$ mag and $\Delta\mathrm{m}$=0.5 mag. Coloured lines show the transient type: (blue) SN Ia, (green) SN Ibc, (red) SN II and (cyan) TDE. Detection efficiency is not 100\% as we account for all sky detection, meaning that a fraction of transients will be obscured by Galactic dust or fall between CCD gaps. }
\label{fig:deteff_z}
\end{figure}

The \gaia  detection efficiency as a transients survey, as mentioned in Section \ref{sec:sims}, is in essence the application of the on-board detection and ground-based candidate selection process on a large number of simulated transients within our mock galaxy catalogue. Provided the number of simulated and detected objects, in this section, we report detection efficiencies for both TDE and SNe.

\subsubsection{TDE detection efficiency}

Fig. \ref{fig:deteff_z} shows the TDE detection efficiency as a function of redshift for $m_{\rm lim}=19$ and $\Delta\mathrm{m}$=0.5. For redshifts closer than 0.02, there is a probability of 0.85 to detect the transients. As we have accounted for a full sky survey with Galactic extinction, there is a small fraction of TDE that will be always obscured by the dust in the Galactic plane. The maximum redshift at which a TDE may be detected, albeit with a detection efficiency of 5\% or lower, is around 0.13 for $m_{\rm lim}=19$ and 0.2 for $m_{\rm lim}=20$.

The survey detection efficiency for PS1 TDE and LR11 TDE, assuming a limiting magnitude of 19 and $\Delta\mathrm{m}$=0.5, is shown in the left panel of Fig. \ref{fig:eff_vs_deltam}. For TDE peaking at 1 mag above the limiting magnitude, a fraction of 0.5 of the candidates will be detected, increasing to nearly 0.7 for transients 2 mag brighter. 

There is not much variation between blue and red galaxies, showing that generally TDE will be bright enough to be detected on top of compact galaxy cores as well as shallower bulges.

\begin{figure*}
\centering
\mbox{
\subfigure{\includegraphics[width=\columnwidth]{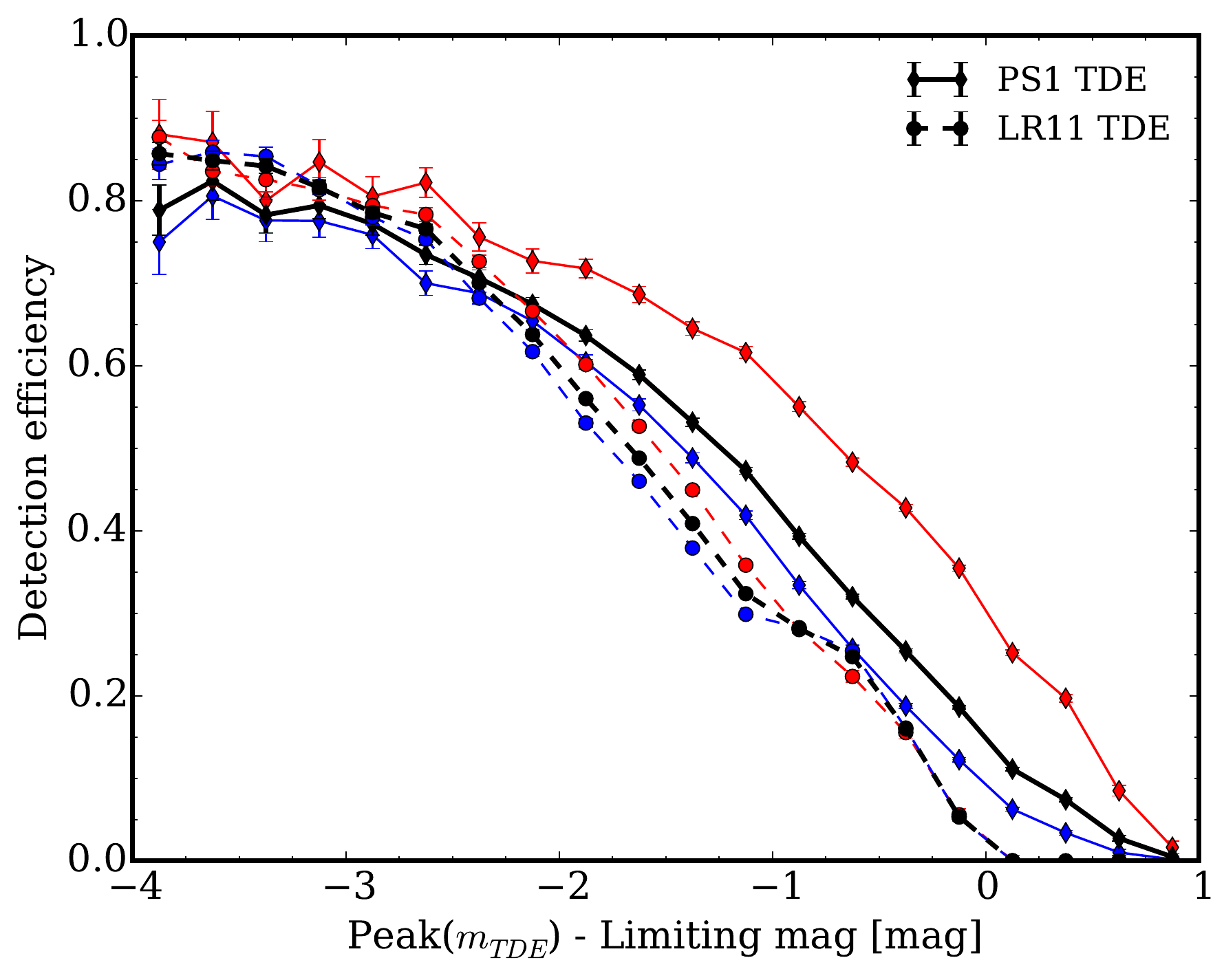} 
\quad
\subfigure{\includegraphics[width=\columnwidth]{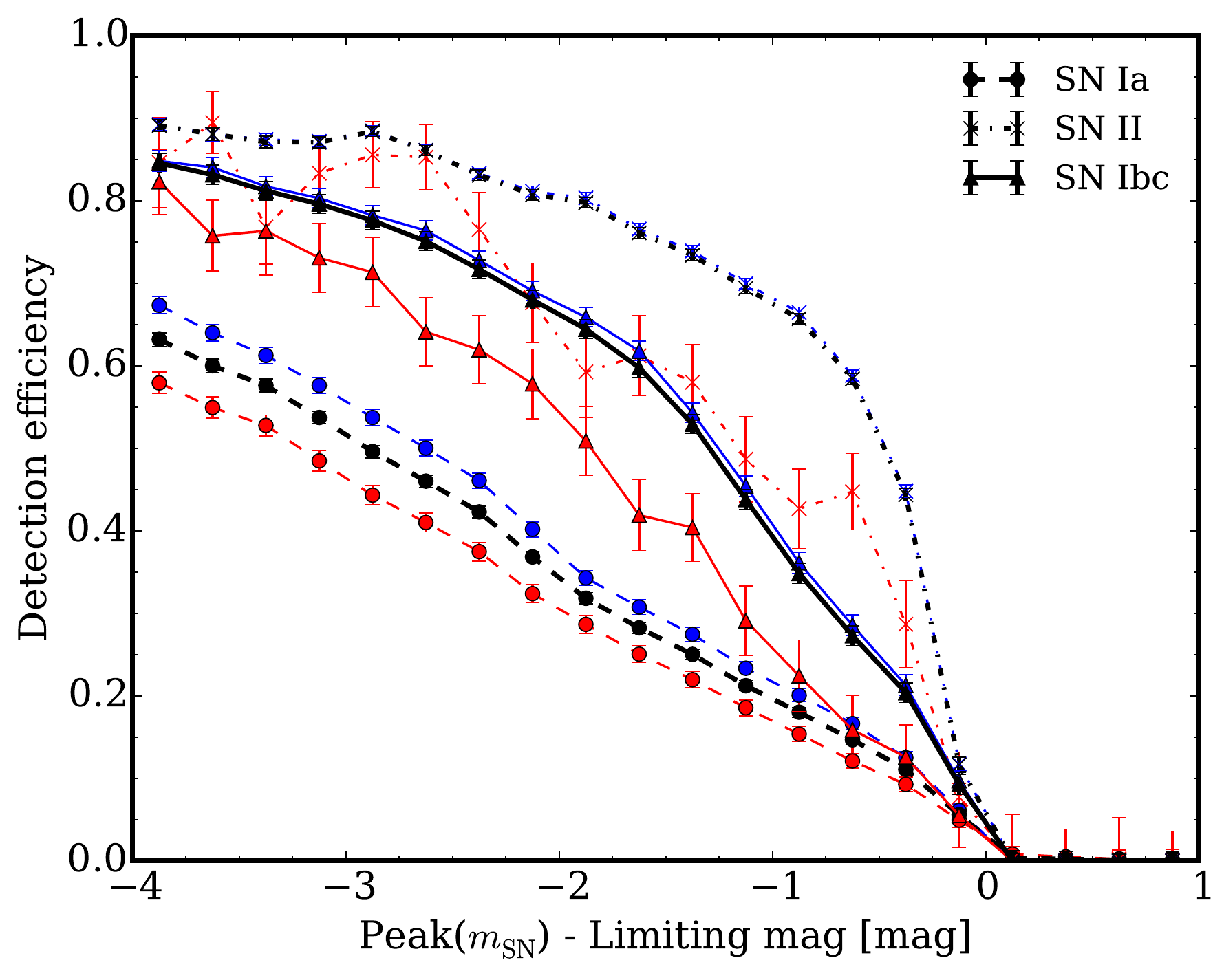} }}}

\caption{Left:\gaia detection efficiency (detected over simulated) as function of the difference between the TDE peak magnitude ($m_{\rm TDE}$) and the limiting magnitude for transient candidates selection,  for $m_{\rm lim}=19$ and $\Delta\mathrm{m}$=0.5 mag. The blue line shows the results for TDE located in the blue galaxy sample, the red line shows the results for TDE in red galaxy sample, and the black line shows the result for both blue and red galaxies. Right: \gaia detection efficiency (detected over simulated) as function of the difference between the SN peak magnitude ($m_{\rm SN}$) and the limiting magnitude for transient candidates selection,  for $m_{\rm lim}=19$ and $\Delta\mathrm{m}$=0.5 mag. The blue line shows the results for SN located in the blue galaxy sample, red line shows the results for SN in the red galaxy sample. The black line shows the result for both blue and red galaxies. Different markers and line styles symbolize the three SN types: SN Ia (dashed), SN Ibc (solid) and SN II (dash-dotted).}
\label{fig:eff_vs_deltam}
\end{figure*}

Fig. \ref{fig:eff_tde_mbh} shows the detection efficiency as a function of the black hole mass for different limiting magnitudes, and assumed light curves: PS1 (top panel) or LR11 (bottom panel). Both results show a dependency on the normalization of the light curve luminosities chosen for PS1 and LR11 models. As explained in Section \ref{sec:method_tde}, observed light curves belong to two PS1 events and we choose between them  according to $M_{\rm BH}$. LR11 light curve models depend on the $M_{\rm BH}$ and their absolute magnitudes are scaled to the luminosity of TDE1 or TDE2 depending as well on $M_{\rm BH}$. The detection efficiency for each bin is computed as the number of detected TDE in the bin over the total number of TDE generate in the same bin that have magnitudes brighter or equal to the limiting magnitude. The results show that for galaxies with lower $M_{\rm BH}$ masses, the detection of TDE is more efficient, because the bulges are less luminous and it is more likely to detect a flare that increases the bulge luminosity more than $\Delta\mathrm{m}$. For galaxies with $M_{\rm BH}$ more massive than 10$^6$ M${\sun}$, the bulges start to be more luminous and dominate over the TDE. However, for higher masses, the scaling of TDE light curves starts approaching the more luminous events (e.g. PS1-10af or TDE2), making these events again easier to detect.

\subsubsection{SN detection efficiency}

The detection efficiency of different types of SNe in \gaia as a function of the difference between the peak SN apparent magnitude and the limiting magnitude for candidate selection is shown in the right panel of Fig. \ref{fig:eff_vs_deltam}. The detection efficiency is shown for red and blue galaxies in our mock galaxy catalogue. Red and blue galaxies generally have different light profiles for the bulge component and different distributions of $B/T$ \citep{LacknerGunn2012}, and the spatial distribution of SNe is different for each type. Our results show that the detection efficiency is higher for blue galaxies, as they generally have smaller and more dispersed bulges. The SNe in late-type galaxies come from younger stellar populations, mostly located on the galactic disc, and therefore at larger distances form the nucleus, which makes them easier to detect as new sources.

Resolved transients \textit{New Source} provide a high accuracy information on the position of the SN. Nuclear transients which are not resolved (\textit{Old Source}) are more challenging to study, as they may originate from different populations: AGN, nuclear SNe or TDE. Ultimately, the shift in the astrometric position of the bulge could possibly discriminate SNe from AGN and TDE. Further work on centroid shift using mission data is needed to quantify this effect. $Orphan$ transients in \gaia, that are discovered without the host galaxy are also problematic: their offset from the galaxy core would need to rely on ground-based coordinates from deeper surveys than \gaia, which generally have less accurate coordinates. Therefore, TDE occurring in less compact bulges will be harder to identify by position only, as TDE will appear as \textit{Orphan} and lack of\gaia accurate astrometry for the position of the host.

\begin{figure}
\centering
\subfigure{\includegraphics[width=\columnwidth]{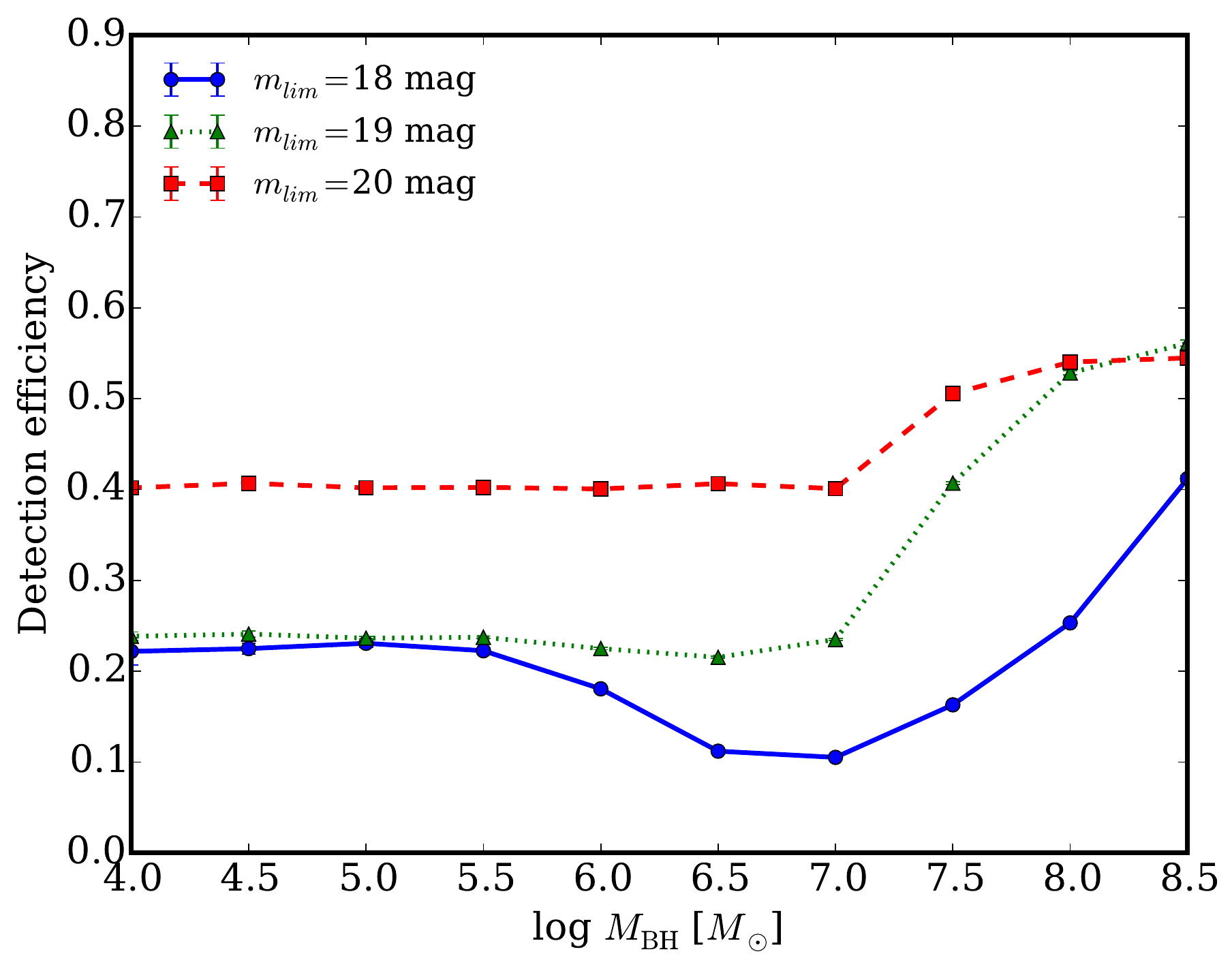} }
\quad \\
\subfigure{\includegraphics[width=\columnwidth]{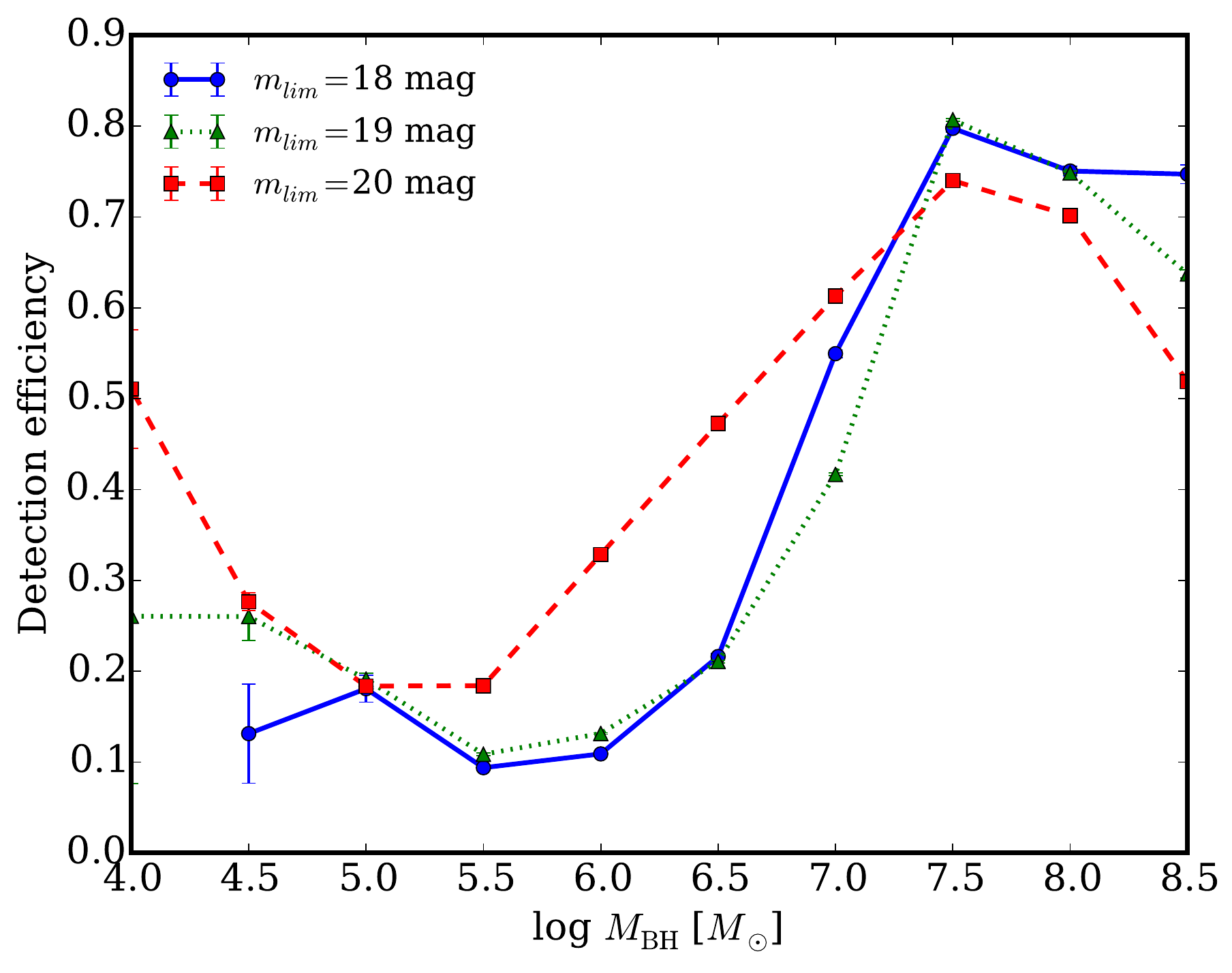} }
\caption{\textit{Upper:} \gaia detection efficiency for TDE with PS1 light curves as function of the host black hole mass, $M_{\rm BH}$. Coloured lines indicate different limiting magnitudes for candidate selection with $\Delta\mathrm{m}$=0.5 mag. The PS1-10jh light curve is used to simulate TDE in less massive galaxies and PS1-10af light curve is used for more massive galaxies. \textit{Lower:} Same for LR11 light curves, which scale with $M_{\rm BH}$. The LR11 light curves are normalized to TDE1 or TDE2 luminosities according to $M_{\rm BH}$. More luminous events are generated in more massive bulges.}
\label{fig:eff_tde_mbh}
\end{figure}

In agreement with the previous results from the \textsc{gibis} on-board detection (Fig. \ref{fig:det_eff_sn}), nuclear unresolved transients (\textit{Old Source} detections) are located at angular separations closer than 0.3-0.4 arcsec. The main SN type detected through this channel is SN Ia, as they are the dominant type in early type galaxies. \textit{Old Source} detections are required to vary more than a certain $\Delta\mathrm{m}$ to be selected as candidates. This is generally achieved by SN Ia, as their intrinsically brighter luminosities can provide a significant flux contribution even on bright bulges, and therefore they are selected as candidates.

\textit{New Source} detections of (mainly) SN Ia show an enhanced detection at closer angular distances when compared to more remote locations. The reason is that SN Ia, being the most luminous class, has a high probability of being brighter than the bulge of its host. As seen in Section \ref{sec:on_board_transient}, the proximity of the bulge provides an additional light contribution, which will make the SNe appear brighter than they really are, allowing them to be selected as candidates for a given limiting magnitude cut.

\textit{Orphan} detections appear for all types of SNe, but they are predominantly CCSNe, which come from a younger stellar populations. The detection efficiency shows a trend that increases for larger angular separations. Provided that the results are based on a limiting magnitude selection, the projected angular distance for closer and brighter SNe is larger than for more distant and fainter ones. Closer SNe also remain longer above the limiting magnitude threshold, so that they are more likely to be sampled by \gaia.

For SN Ia, the central plot reveals the existence of a `blind spot' around 0.2--0.3 arcsec for transients with an identified host in \gaia. These transients are too faint to dominate the emission, and too close to be resolved as independent detections. These objects, when detected, will appear as a mixture of blended and resolved points, depending on the scanning angle.

\subsection{Rate prediction} \label{sec:rate}

\subsubsection{TDE detection rate prediction}

The number of TDE detectable by \gaia is computed using MC simulations, which provide the basis for Equation \ref{eq:ntde}. For the candidate selection process, we assumed different limiting magnitudes for candidate selection (18, 19 and 20) and two thresholds for the increase in the bulge magnitude (0.3 and 0.5). Table \ref{tab:tde_det_eff} provides an overview of the expected number of detections. 

Varying $\Delta\mathrm{m}$ has a minimal impact on the predicted TDE detection rate. This means that this magnitude threshold for selecting candidates does not play a major role in the final number of TDE. Intrinsically, TDE are very bright events which are point sources. In Section \ref{sec:on_board_transient} we show that they will be detected even if their luminosity is 1 magnitude below that of their host galaxy bulge.

Finally, we observe a small difference between the number of TDEs obtained with different assumptions for the light curves. The PS1 lightcuves do not scale with $M_{\rm BH}$ and therefore predict both longer decay rates and brighter flares for even the smallest population of SMBH. This has the implication that if we assume the same light curve characteristics for all SMBH masses,
more of these objects will be detected with the \gaia cadence. The LR11 light curves peak brightness and time-scales depend on the mass of the SMBH, and therefore predict shorter events for the low mass end of the SMBH population.

The reported numbers of detections may seem high when compared to SNe, as the intrinsic ratio of TDE to SNe is about 1/100. However, the ratio of detections is higher, because TDE have a higher detection efficiency, related to their higher luminosity and longer duration.

\begin{figure*}
\centering
\mbox{
\subfigure{\includegraphics[width=\columnwidth]{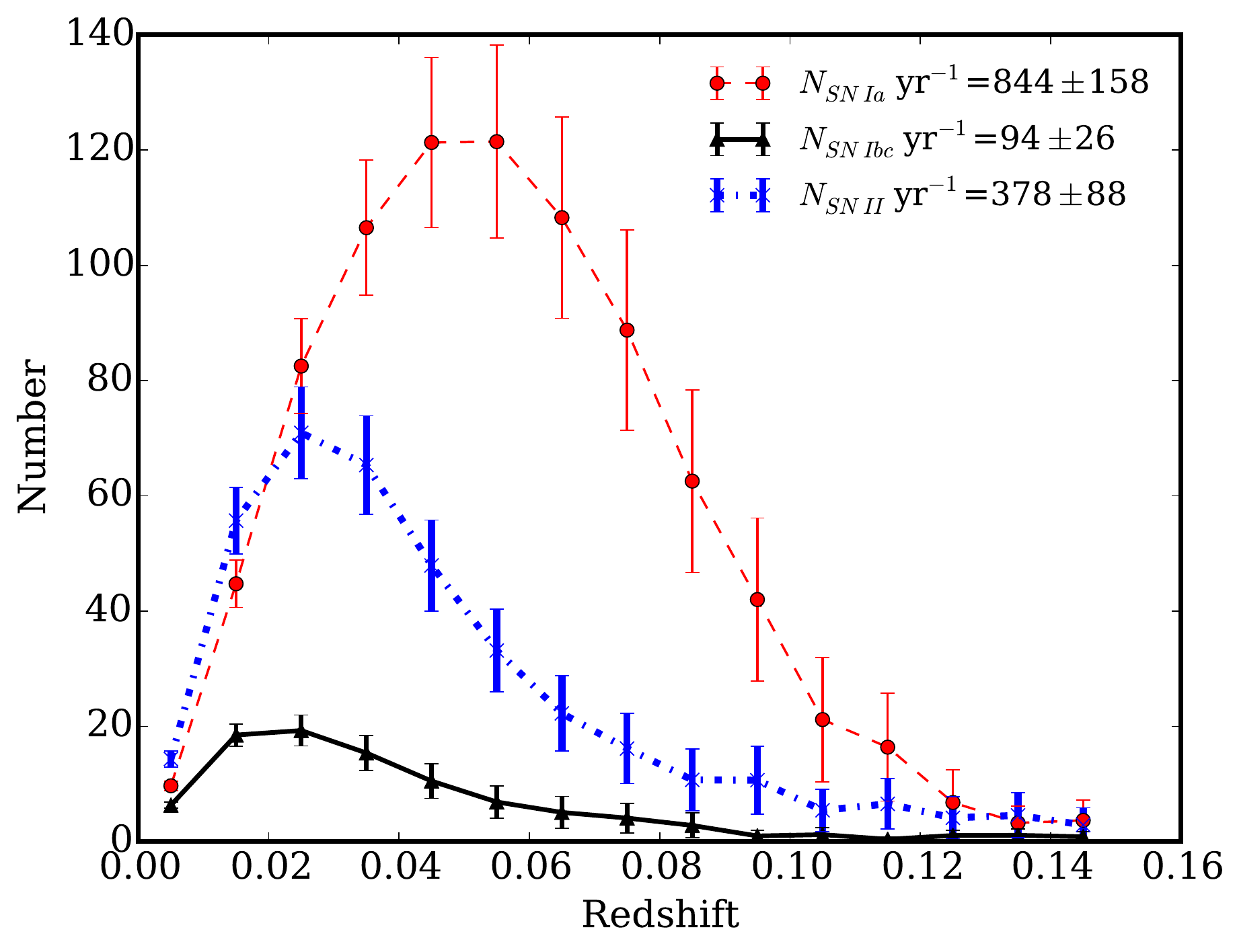} 
\quad
\subfigure{\includegraphics[width=\columnwidth]{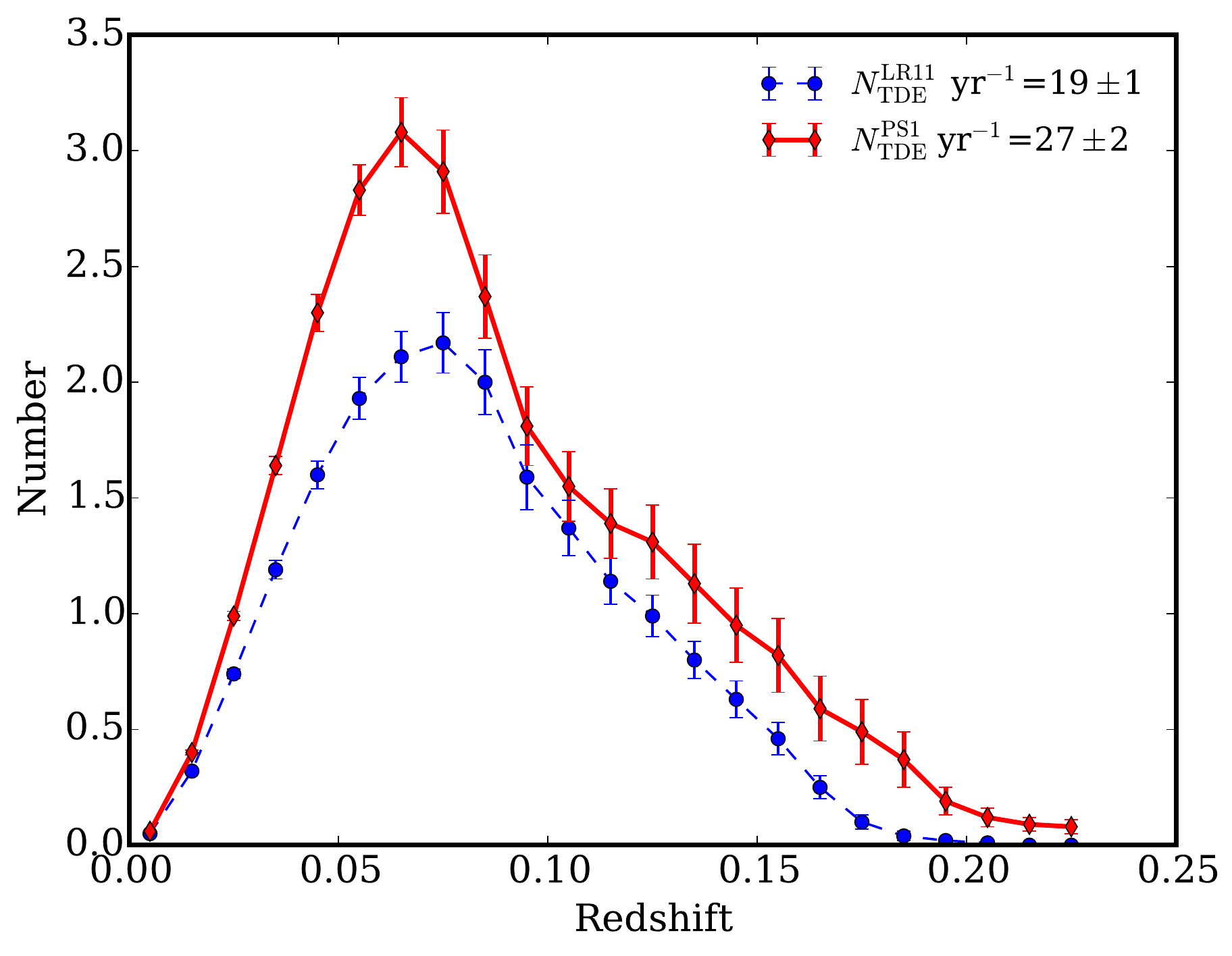} }} }
\caption{{\it Left}: Expected numbers of SNe for each redshift bin of 0.01. The results are shown for a limiting magnitude for transient candidates of $m_{\rm lim}=19$ mag, and flux increase of $\Delta\mathrm{m}$=0.5 mag.The scatter in each bin includes the uncertainty on the number of objects per bin, error on the SN rate and error on the galaxy luminosity function. {\it Right}:  Expected numbers of TDE using same selection criteria: $m_{\rm lim}=19$ mag, and flux increase of $\Delta\mathrm{m}$=0.5 mag. The two curves show the numbers of TDE assuming observed light curves from PS1 or the model light curves from LR11. }
\label{fig:sn_tde_numbers}
\end{figure*}

\begin{table}
 \centering
 \begin{minipage}{80mm}
 \centering
  \caption{Expected number of detections for different candidate selection criteria for limiting magnitude ($m_{\rm lim}$) and minimum increase in the host bulge magnitude $\Delta\mathrm{m}$. Results are given for TDE with light curves from observed events (PS1) and models (LR11).}
  \begin{tabular}{@{}llllll@{}}
  \hline
 $m_{\rm lim}$ & $\Delta\mathrm{m}$ & $N_{\rm TDE}^{\rm PS1}$  & $N_{\rm TDE}^{\rm LR11}$  \\ 

(mag) & (mag)  & (yr$^{-1}$) & (yr$^{-1}$)  			\\ \hline
	18 & 0.5 	& 7  $\pm$ 1 		& 4 $\pm$ 1 		\\
	18 & 0.3 	&  7  $\pm$ 1 		& 4 $\pm$ 1 		\\ \hline
	19 & 0.5 	&  27  $\pm$ 2 		& 19 $\pm$ 1		\\
	19 & 0.3 	&  27  $\pm$ 2 		& 20 $\pm$ 21	\\ \hline
	20 & 0.5 	&  76   $\pm$ 5		& 51 $\pm$ 6 	\\
    20 & 0.3 	& 75  $\pm$ 5 		& 53 $\pm$ 6	\\ \hline
    \label{tab:tde_det_eff}
\end{tabular}
\end{minipage}
\end{table}

\subsubsection{SN detection rate prediction}

The prediction for the number of SNe to be detected by \gaia is computed according to the approach described in \ref{sec:method_sn}. The results for the MC simulations and different candidate selection processes parametrized by  $m_{\rm lim}$ and $\Delta\mathrm{m}$ are shown in Table \ref{tab:sn_det_numbers}. We provide estimates for the unresolved, resolved and \textit{Orphan} transients, and an estimation of the number of transients within the central kpc of their host galaxies. Fig. \ref{fig:sn_tde_numbers} shows the distribution for the expected number over different redshift bins. For the most likely candidate selection parameters of $m_{\rm lim}=19$ mag and $\Delta\mathrm{m} = 0.5$ mag, we observe approximately 850 SN Ia, 100 SN Ibc and 380 SN type II. The numbers per detection method: \textit{New Source}, \textit{Old Source} or \textit{Orphan}, show that all CCSNe will be located in galaxies non-detectable by \gaia. For SN Ia, about 5\% will be \textit{Old Sources}, $\sim$20\% will be \textit{New Sources}, and the remaining 75\% will be \textit{Orphan}; they will occur in galaxies that would not be detected by \gaia. For SN Ia, we would expect ~20\% of them to be detected within the central kpc within their hosts. This number drops for CCSNe, because of their less compact radial distribution. We would expect only $\sim$3\% of them to be located in the central kpc of their host galaxies. The low number of SN II in the centre is due to the low number statistics of such events in the core, and the difficulty in detecting such events using the \textit{Old Source} method, as their absolute magnitudes are approximately 1--3 mag fainter than for SN Ia. 

The epoch of discovery for different types of SN are shown in Fig. \ref{fig:sn_dis_epoch}, assuming a limiting magnitude of 19 for the selection of transient candidates. In agreement with \cite{BelokurovEvans2003}, about one third of SN Ia will be detected before maximum light. For SN Ibc, which have longer raise times, around 50\% are predicted to be detected before maximum. The majority of detections for SN type I will be within 20 d post maximum light. For SN type II, around 40\% will be younger than 2 weeks when detected by \gaia. Assuming a fainter limiting magnitude for the candidate selection, the fraction of SNe detected at early epochs increases. These SNe, located in the faint end, will be the majority of the alerts. As the visible part of their light curves will be reduced to the area around maximum brightness only, they will either be detected around the peak, or will never be detected, because they will be too faint if scanned at later epochs.

\begin{figure*}
\centering
\includegraphics[width=2.\columnwidth]{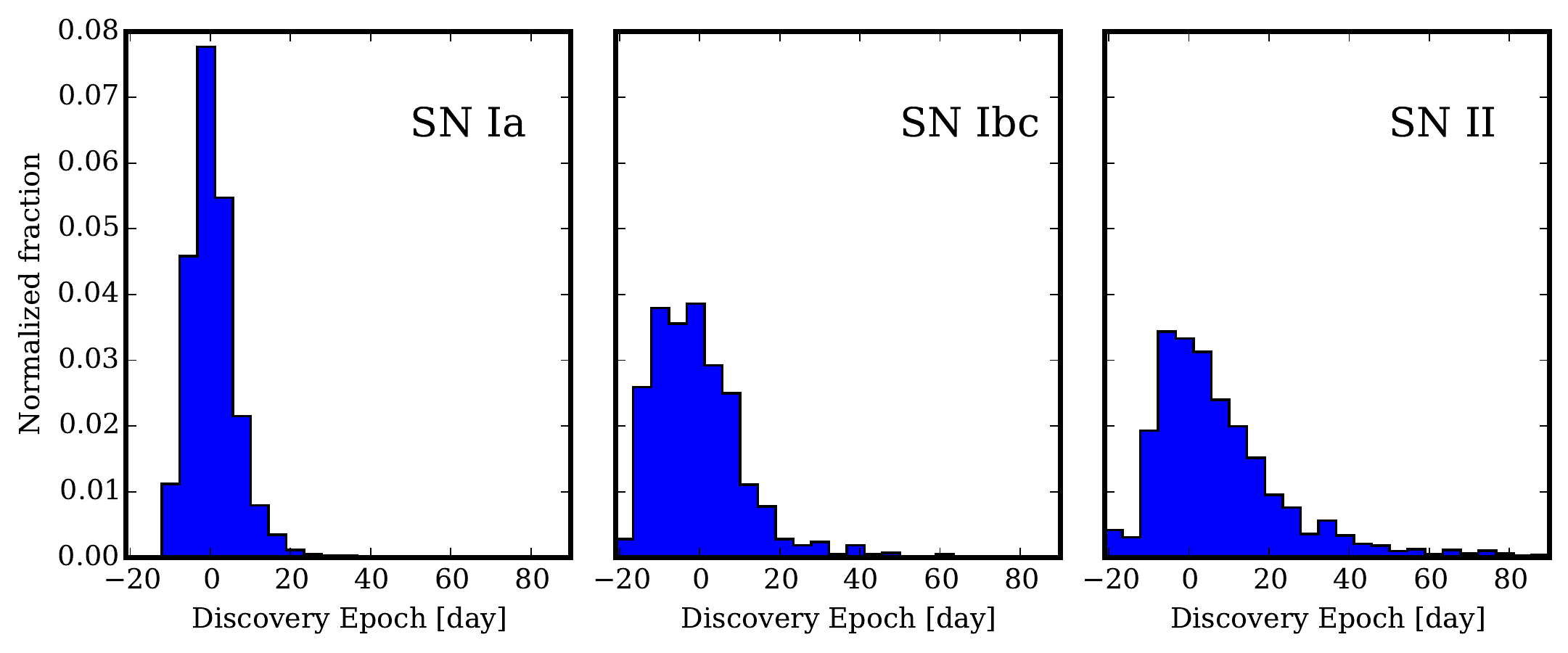} 

\caption{Normalized fraction of objects for a given discovery epoch.  The results are shown for a limiting magnitude for transient candidates of $m_{\rm lim}=19$, and flux increase of $\Delta\mathrm{m}$=0.5 mag. Epoch zero refers to the day of maximum brightness in visual $V$ band.}
\label{fig:sn_dis_epoch}
\end{figure*}

\begin{table*}
 \centering
 %\begin{minipage}{140mm}
  \caption{ Predicted numbers of SNe depending on the type, the limiting candidate magnitude, $m_{\rm lim}$, the minimum magnitude variation for existing objects $\Delta\mathrm{m}$ and detection method. \textit{Old Source} detections are not resolved from the host galaxy and the candidates are selected by the change in magnitude of the source. \textit{New Source} detections are new transients resolved from the host galaxy and the host is detected as well. \textit{Orphan} transients do not have their hosts detected by \gaia, although they may exist in other surveys. An additional column indicates what number of detected sources will occur within the central kpc in their host galaxies.}
  \begin{tabular}{@{}lllcccccc@{}}
  \hline
SN type & $m_{\rm lim}$ & $\Delta\mathrm{m}$  & Total &  \textit{Old Source} & \textit{New Source} & \textit{Orphan}  & Distance $<$1 kpc \\ 
 & (mag) & (mag)  & ( yr$^{-1}$) & (yr$^{-1}$) & (yr$^{-1}$)&(yr$^{-1}$) & (yr$^{-1}$) \\ \hline

SN Ia & 17 & 0.3 & 67$\pm$9   & 1 $\pm$ 1 & 14 $\pm$ 2 & 49 $\pm$ 7 & 14 $\pm$ 3\\
SN Ibc & 17 & 0.3 & 8$\pm$1   & 0 $\pm$ 0 & 0 $\pm$ 0 & 8 $\pm$ 1 & 0 $\pm$ 0\\
SN II & 17 & 0.3 & 33$\pm$8   & 0 $\pm$ 0 & 0 $\pm$ 0 & 33 $\pm$ 8 & 1 $\pm$ 0\\ \hline
SN Ia & 17 & 0.5 & 66$\pm$9   & 1 $\pm$ 1 & 14 $\pm$ 2 & 49 $\pm$ 7 & 14 $\pm$ 3\\
SN Ibc & 17 & 0.5 & 8$\pm$1   & 0 $\pm$ 0 & 0 $\pm$ 0 & 8 $\pm$ 1 & 0 $\pm$ 0\\
SN II & 17 & 0.5 & 33$\pm$8   & 0 $\pm$ 0 & 0 $\pm$ 0 & 33 $\pm$ 8 & 1 $\pm$ 0\\ \hline
SN Ia & 18 & 0.3 & 250$\pm$36   & 16 $\pm$ 7 & 60 $\pm$ 10 & 170 $\pm$ 24 & 54 $\pm$ 15\\
SN Ibc & 18 & 0.3 & 26$\pm$5   & 0 $\pm$ 0 & 0 $\pm$ 0 & 25 $\pm$ 5 & 1 $\pm$ 0\\
SN II & 18 & 0.3 & 108$\pm$25   & 0 $\pm$ 0 & 1 $\pm$ 0 & 107 $\pm$ 24 & 4 $\pm$ 1\\ \hline
SN Ia & 18 & 0.5 & 246$\pm$35   & 12 $\pm$ 6 & 60 $\pm$ 10 & 170 $\pm$ 24 & 51 $\pm$ 14\\
SN Ibc & 18 & 0.5 & 26$\pm$5   & 0 $\pm$ 0 & 0 $\pm$ 0 & 25 $\pm$ 5 & 1 $\pm$ 0\\
SN II & 18 & 0.5 & 108$\pm$25   & 0 $\pm$ 0 & 1 $\pm$ 0 & 107 $\pm$ 24 & 4 $\pm$ 1\\ \hline
SN Ia & 19 & 0.3 & 889$\pm$135   & 74 $\pm$ 30 & 164 $\pm$ 32 & 636 $\pm$ 96 & 202 $\pm$ 58\\
SN Ibc & 19 & 0.3 & 97$\pm$22   & 0 $\pm$ 0 & 1 $\pm$ 1 & 94 $\pm$ 21 & 3 $\pm$ 1\\
SN II & 19 & 0.3 & 381$\pm$81   & 0 $\pm$ 0 & 1 $\pm$ 1 & 380 $\pm$ 81 & 12 $\pm$ 5\\ \hline

SN Ia & 19 & 0.5 & 844$\pm$158   & 44 $\pm$ 20 & 164 $\pm$ 32 & 636 $\pm$ 96 & 170 $\pm$ 49\\
SN Ibc & 19 & 0.5 & 94$\pm$26   & 0 $\pm$ 0 & 1 $\pm$ 1 & 94 $\pm$ 21 & 3 $\pm$ 1\\
SN II & 19 & 0.5 & 378$\pm$88   & 0 $\pm$ 0 & 1 $\pm$ 1 & 380 $\pm$ 81 & 12 $\pm$ 5\\ \hline

SN Ia & 20 & 0.3 & 2990$\pm$419   & 93 $\pm$ 38 & 297 $\pm$ 67 & 2575 $\pm$ 382 & 555 $\pm$ 149\\
SN Ibc & 20 & 0.3 & 350$\pm$71   & 0 $\pm$ 0 & 2 $\pm$ 2 & 345 $\pm$ 69 & 10 $\pm$ 6\\
SN II & 20 & 0.3 & 1330$\pm$225   & 0 $\pm$ 0 & 3 $\pm$ 2 & 1328 $\pm$ 226 & 41 $\pm$ 18\\ \hline
SN Ia & 20 & 0.5 & 2950$\pm$414   & 54 $\pm$ 26 & 297 $\pm$ 67 & 2575 $\pm$ 382 & 514 $\pm$ 139\\
SN Ibc & 20 & 0.5 & 350$\pm$71   & 0 $\pm$ 0 & 2 $\pm$ 2 & 345 $\pm$ 69 & 10 $\pm$ 6\\
SN II & 20 & 0.5 & 1330$\pm$225   & 0 $\pm$ 0 & 3 $\pm$ 2 & 1328 $\pm$ 226 & 41 $\pm$ 18\\ \hline

    \label{tab:sn_det_numbers}
\end{tabular}
%\end{minipage}
\end{table*}

In order to assess the impact of the \gaia detection process on transient recovery, Fig. \ref{fig:sn_gal_radi} shows the simulated and the recovered surface density for different types of SNe as function of their relative radial distance, $\gamma=R_{\rm SN}/h$, where $R_{\rm SN}$ is the SN angular separation from the host and $h$ is the galaxy disc scale length. To evaluate the recovery rate, we computed both the simulated and detected SN surface densities for different relative radial distances as $\Sigma=N_i/[ \mathrm{\pi} (\gamma_i - \gamma_{i-1})]$, where $N_i$ is the number of SNe in bin $i$, between the relative distance $\gamma_i$ and $\gamma_{i-1}$. Both the simulated and the retrieved distribution have been normalized to 1 to allow easier comparison. The figure shows that there are no noticeable biases related to lower detection of SNe in central regions. On contrary, we see an enhancement of the number of detected SN Ia for relative distances closer than 0.1. As mentioned, the contribution of the bulge light may enhance the intrinsic brightness of these type of objects beyond $m_{\rm lim}$, allowing them to be selected as candidates.

\begin{figure*}
\centering
\includegraphics[width=2.0\columnwidth]{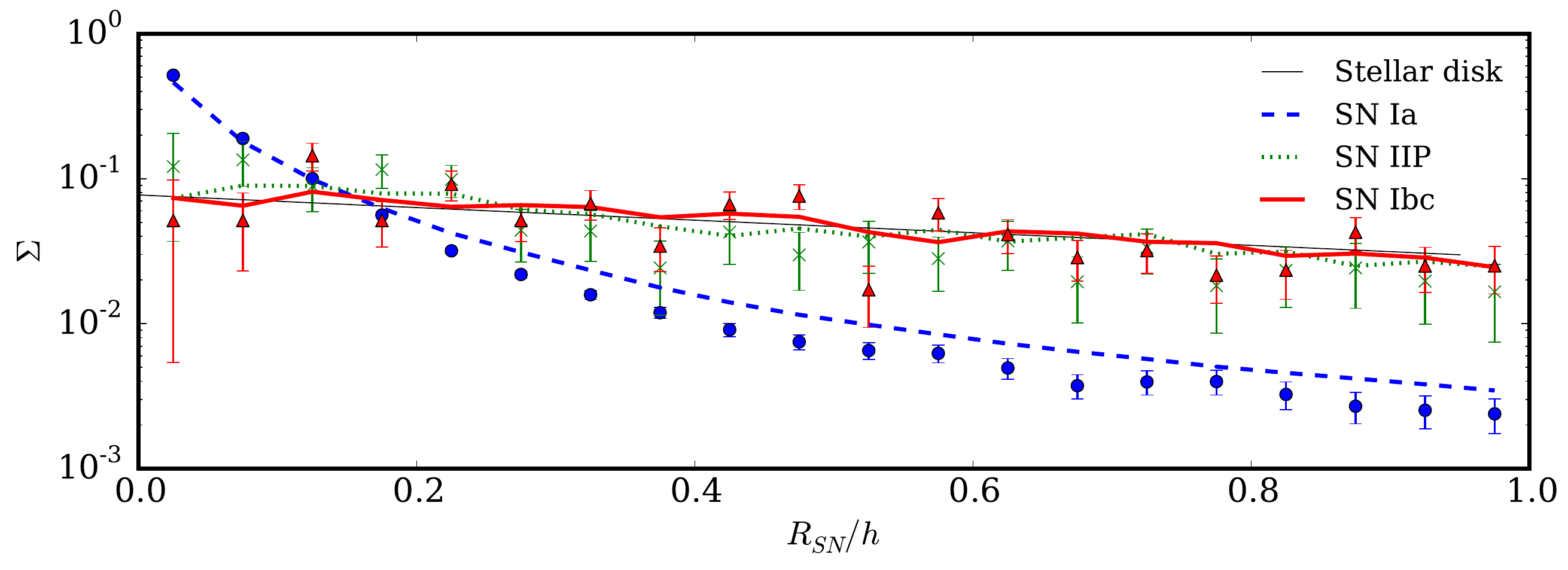} 

\caption{Normalized surface density of different types of SNe for the generated distribution with peak magnitude of 19 or brighter (lines) and the detected surface density for the same SN type (symbols). $R_{\rm SN}$ represents the radial distance of the SN and the $h$ is the exponential disc scale length. The black thin line represents the light distribution in an exponential disc, used to generate the CCSNe population. }
\label{fig:sn_gal_radi}
\end{figure*}

\section{Discussion} \label{sec:discussion}

\subsection{Comparison with literature} \label{sec:comparison}

Our prediction for the numbers of all types of SNe detected with \gaia agree with the numbers reported in previous works \citep{BelokurovEvans2003, Altavilla2012}. \cite{BelokurovEvans2003} also assumed SN rates to be a function of galaxy type and luminosity. There is a natural scatter in the results due to the slightly different rates that we adopted for SNe and differences in the generated light curves. For a limiting magnitude of 19, their work predicts approximately 610 SNe Ia and 690 CCSNe per year. We obtain a slightly larger number for SNe Ia. This is due to the fact that these intrinsically bright transients will be effectively detected in both red and blue galaxies, and, as seen, the bulge light contribution may even enhance their detection. The number of CCSNe, which are intrinsically fainter than SN Ia, is lower in our predictions. Our results predict $\sim$100 CCSNe less than the \cite{BelokurovEvans2003} study. This is explained by the loss of this intrinsically fainter SN class when they occur in redder galaxies, which have more prominent bulges, as shown in Fig.s \ref{fig:eff_vs_deltam}.

Comparison with the predicted numbers by \cite{Altavilla2012}, based on volumetric SN rates, shows a noticeable difference for CCSNe as well. Imposing a limiting magnitude of 19 for the selection of candidates, the predicted number of SNe Ia is about 1070. This is consistent with our estimates within the numerical uncertainties. However, for CCSNe, they estimate $\sim$190 detections per year, but we predict more than twice this number and located at larger distances. In our predictions, given that we adopted a scatter of 1.37 magnitudes for the absolute magnitude of SNe II, we do not expect a sharp cut in the detection at redshift $z=$0.04, but rather a more extended wing of detected transients towards higher redshifts, up to 0.14, according to Fig. \ref{fig:sn_tde_numbers}.

The main conclusion from these comparisons is that differences in the predicted numbers exist, but they are within the statistical and systematic errors for SN luminosity functions and rates in the literature, as well as the assumptions made on the light curves used to predict the brightness of SNe.

\subsection{Caveats} \label{sec:caveats}

The results reported in our study are sensitive to the limitation of the \textsc{gibis} simulator when modelling the structure of galaxies, as well as by the methodology we used to create the mock galaxy catalogue. In this section we will discuss the possible implications of an assumption of a simple dichotomy of bulge light profiles; De Vaucouleurs' profile and exponential profile.

In this work we assumed that all early-type galaxies will have a more compact light profile for the bulge, while late-type galaxies will have much flatter light distributions. The direct effect is that our population of SNe type Ia, which arise in early-type galaxies, is positioned at smaller offsets from the nucleus of their hosts, following a more peaky light distribution. Therefore, we will expect our simulations to produce an enhanced number of SNe Ia in the central regions of galaxies when compared to a real sample. On the contrary, CCSNe are simulated following disc light only, meaning that we do not include an enhancement of the nuclear CCSN population in starburst galaxies or LIRGs, as e.g. suggested by \cite{Herrero-Illana2012}. Therefore, we expect a lower number of CCSNe in the central regions compared with observations.

Finally, the extinction distribution in the host galaxies is also an important source of uncertainty, especially for CCSNe located in nearby actively star forming galaxies (such as LIRGs). The \gaia sample of CCSNe, along with ground-based follow-up, will also provide a valuable dataset to study the host galaxy extinction in the nuclear and circumnuclear regimes by comparison between the detected and expected numbers of SNe.

Another caveat is the uncertainty on the final parameters to be adopted by the VPU on board \gaia \citep{DeBruijne2015}. This work is based on the original VPU parameters available through \textsc{gibis}. However, the current version of the VPU running on board has slightly updated parameter values, which result in an increased detection rate for the brightest and smallest spiral galaxies. The effect of these new parameters over the originally adopted ones is a higher rate of objects with a detected host in \gaia and a lower number of hostless transients. However, given the galaxy population in our Universe Model, we assess that the number of galaxies falling within the new detectable parameter space is very low (less than 0.5\%). Therefore, we conclude that no major differences are to be expected in the results presented here.

The new VPU parameters and increased downlink rate have contributed to increase the \gaia mission limiting magnitude to approximately 20.7. Although more transients will be detected in this faint end, their observations are excluded from the search and analysis of new transient candidates. Processing the very faint end represents a considerable demand in computing time and the rate of true positives over false positives in this low S/N regime is expected to be very low.

\section{Conclusions and Summary} \label{sec:conclusions}

\gaia has a unique capability for identifying transients in the nuclear regime (angular separation from nucleus $<$ 1 arcsec). In this paper, we provide a quantitative analysis of this capability. We compute the detection efficiency of \gaia for transients as a function of their magnitude and angular separation from the nucleus of their host galaxies. We calculate the fraction of transients that we would expect to find in different regimes:  non-resolved from the host, resolved from the host and with no host in \gaia data. We include the effect of the \gaia NSL with an average sampling cadence of 30 d. The main conclusions of our analysis are listed below:
 
\begin{enumerate}
 \item The limiting angular separation for \gaia to resolve an SN from the bulge of its host galaxy and generate a new detection is dependent on the scanning angle and the magnitude difference between the two sources. For galaxies and SNe fainter than 16 magnitude, the minimum distance to resolve a galaxy bulge from the SN with a probability of 50\% or larger, is around 0.2 arcsec.
 \item Transients occurring in detected galaxies, if not resolved, will be detected as an increase in flux of an already identified source. Detection of a shift in the position of the centroid ($>$ 0.01 arcsec) will provide additional information to constrain the nuclear location of the transients. 
 \item Imposing a limiting \gaia magnitude of 19 for transient candidate selection, the estimated number of detected SNe is around 1300 per year. About $\sim$15\% of them are expected to be localized at nuclear offsets smaller than 1 arcsec, (or 2 kpc for redshifts z$\leq$0.1).
\item  \gaia is expected to detect around 20--30 TDE each year for a limiting magnitude of $G$=19.
\item The expected number of unresolved SNe, is $\sim 40$ per year, which is of the same order of magnitude as the number of TDE. In order to distinguish these two classes of transients, further work is need to be able to measure the shift in the centroid for unresolved detections so as to reduce the number of SNe that could be mistaken as TDE.
\item \textit{Orphan} SNe are expected to be the most common type of transient in \gaia, given the characteristics of galaxy detectability. About $\sim$70\% of SN Ia and $\sim$90\% of CCSNe in \gaia are going to be detected as hostless events. However, hostless in \gaia does not imply hostless in other ancillary catalogues.
\item Although the \gaia detection efficiency for CCSNe appears to be slightly lower in the central regions ($<$0.3 arcsec) of galaxies with bright bulges, the overall effect on the normalized distance distribution is minimal, allowing us to overcome the so-called `Shaw effect' with the SN sample obtained by \gaia.
\end{enumerate}

The future development of the work presented in this paper, is the analysis of the detection efficiency of transients with \gaia mission data, in comparison with ground-based surveys. Further into the mission, fast stacking algorithms for the reconstruction of 2D profiles from \gaia 1D  scans \citep{Harrison2011} will be used to identify the galaxy morphology of SN hosts using \gaia only \citep{Krone-Martins2013}. This will allow us to expand the detection efficiency results presented here into a complete understanding of transient detection for the real sample of galaxies, allowing us to derive final mission detection efficiencies and rates for the nearby extragalactic transient population.

\section*{Acknowledgements}

We thank Morgan Fraser, Jos de Bruijne, Samy Azaz and Alberto Krone-Martins for useful comments and discussions.
The research leading to these results has received funding from the European Union Seventh Framework Programme ([FP7/2007-2013]) under grant agreement num. 264895. SvV was supported by NASA through Hubble Fellowship grant HST-HF2-51350.001-A awarded by the Space Telescope Science Institute, which is operated by the Association of Universities for Research in Astronomy, Inc., for NASA, under contract NAS5-26555. 
LW acknowledges support from the Polish NCN `Harmonia' grant No. 2012/06/M/ST9/00172.
Simulated data provided by the Simulation Unit (CU2) of the \gaia Data Processing Analysis Consortium (DPAC) have been used to complete this work. The simulations have been done at CNES (Centre national d'\'{e}tudes spatiales). They are gratefully acknowledged for this contribution.

\label{lastpage}

\bibliographystyle{mn2e}
\bibliography{references}

\begin{thebibliography}{57}
\expandafter\ifx\csname natexlab\endcsname\relax\def\natexlab#1{#1}\fi

\bibitem[{{Altavilla} {et~al}\mbox{.}(2012){Altavilla}, {Botticella},
  {Cappellaro}, \& {Turatto}}]{Altavilla2012}
{Altavilla} G., {Botticella} M.~T., {Cappellaro} E., {Turatto} M., 2012, \apss,
  341, 163

\bibitem[{{Arcavi} {et~al}\mbox{.}(2014){Arcavi}, {Gal-Yam}, {Sullivan}, {Pan},
  {Cenko}, {Horesh}, {Ofek}, {De Cia}, {Yan}, {Yang}, {Howell}, {Tal},
  {Kulkarni}, {Tendulkar}, {Tang}, {Xu}, {Sternberg}, {Cohen}, {Bloom},
  {Nugent}, {Kasliwal}, {Perley}, {Quimby}, {Miller}, {Theissen}, \&
  {Laher}}]{Arcavi2014}
{Arcavi} I. {et~al.}, 2014, \apj, 793, 38

\bibitem[{{Babusiaux}(2005)}]{babusiaux05}
{Babusiaux} C., 2005, in ESA Special Publication, Vol. 576, The
  Three-Dimensional Universe with Gaia, {C.~Turon, K.~S.~O'Flaherty, \&
  M.~A.~C.~Perryman}, ed., pp. 417--+

\bibitem[{{Babusiaux} {et~al}\mbox{.}(2013){Babusiaux}, {Sartoretti},
  {Leclerc}, {Chereau}, \& {Weller}}]{babusiaux13}
{Babusiaux} C., {Sartoretti} P., {Leclerc} N., {Chereau} F., {Weller} M., 2013,
  {The Gaia Instrument and Basic Image Simulator GIBIS 13.0 - User Guide ({\it
  GAIA-C2-SP-OPM-CB-003-14-1}) }. {http://gibis.cnes.fr/}

\bibitem[{{Baldry} {et~al}\mbox{.}(2004){Baldry}, {Glazebrook}, {Brinkmann},
  {Ivezi{\'c}}, {Lupton}, {Nichol}, \& {Szalay}}]{Baldry2004}
{Baldry} I.~K., {Glazebrook} K., {Brinkmann} J., {Ivezi{\'c}} {\v Z}., {Lupton}
  R.~H., {Nichol} R.~C., {Szalay} A.~S., 2004, \apj, 600, 681

\bibitem[{{Belokurov} \& {Evans}(2003)}]{BelokurovEvans2003}
{Belokurov} V.~A., {Evans} N.~W., 2003, \mnras, 341, 569

\bibitem[{{Blagorodnova} {et~al}\mbox{.}(2014){Blagorodnova}, {Koposov},
  {Wyrzykowski}, {Irwin}, \& {Walton}}]{Blagorodnova2014}
{Blagorodnova} N., {Koposov} S.~E., {Wyrzykowski} {\L}., {Irwin} M., {Walton}
  N.~A., 2014, \mnras, 442, 327

\bibitem[{{Bonifacio}, {Monai} \& {Beers}(2000){Bonifacio}, {Monai}, \&
  {Beers}}]{Bonifacio2000}
{Bonifacio} P., {Monai} S., {Beers} T.~C., 2000, \aj, 120, 2065

\bibitem[{{Cappellaro} {et~al}\mbox{.}(1997){Cappellaro}, {Turatto},
  {Tsvetkov}, {Bartunov}, {Pollas}, {Evans}, \& {Hamuy}}]{Cappellaro1997}
{Cappellaro} E., {Turatto} M., {Tsvetkov} D.~Y., {Bartunov} O.~S., {Pollas} C.,
  {Evans} R., {Hamuy} M., 1997, \aap, 322, 431

\bibitem[{{Chilingarian}, {Melchior} \& {Zolotukhin}(2010){Chilingarian},
  {Melchior}, \& {Zolotukhin}}]{Chilingarian2010}
{Chilingarian} I.~V., {Melchior} A.-L., {Zolotukhin} I.~Y., 2010, \mnras, 405,
  1409

\bibitem[{{Chilingarian} \& {Zolotukhin}(2012)}]{Chilingarian2012}
{Chilingarian} I.~V., {Zolotukhin} I.~Y., 2012, \mnras, 419, 1727

\bibitem[{{Chornock} {et~al}\mbox{.}(2014){Chornock}, {Berger}, {Gezari},
  {Zauderer}, {Rest}, {Chomiuk}, {Kamble}, {Soderberg}, {Czekala}, {Dittmann},
  {Drout}, {Foley}, {Fong}, {Huber}, {Kirshner}, {Lawrence}, {Lunnan}, \&
  {Marion}}]{Chornock2014}
{Chornock} R. {et~al.}, 2014, \apj, 780, 44

\bibitem[{{Dahlen} {et~al}\mbox{.}(2012){Dahlen}, {Strolger}, {Riess},
  {Mattila}, {Kankare}, \& {Mobasher}}]{Dahlen2012}
{Dahlen} T., {Strolger} L.-G., {Riess} A.~G., {Mattila} S., {Kankare} E.,
  {Mobasher} B., 2012, \apj, 757, 70

\bibitem[{{de Bruijne}(2012)}]{DeBruijne2012}
{de Bruijne} J.~H.~J., 2012, \apss, 341, 31

\bibitem[{{de Bruijne} {et~al}\mbox{.}(2015){de Bruijne}, {Allen}, {Azaz},
  {Krone-Martins}, {Prod'homme}, \& {Hestroffer}}]{DeBruijne2015}
{de Bruijne} J.~H.~J., {Allen} M., {Azaz} S., {Krone-Martins} A., {Prod'homme}
  T., {Hestroffer} D., 2015, \aap, 576, A74

\bibitem[{{de Souza} {et~al}\mbox{.}(2014){de Souza}, {Krone-Martins}, {dos
  Anjos}, {Ducourant}, \& {Teixeira}}]{deSouza2014}
{de Souza} R.~E., {Krone-Martins} A., {dos Anjos} S., {Ducourant} C.,
  {Teixeira} R., 2014, \aap, 568, A124

\bibitem[{{de Vaucouleurs}(1948)}]{deVaucouleurs1948}
{de Vaucouleurs} G., 1948, Annales d'Astrophysique, 11, 247

\bibitem[{{Filippenko}(1997)}]{Filippenko1997}
{Filippenko} A.~V., 1997, \araa, 35, 309

\bibitem[{{F{\"o}rster} \& {Schawinski}(2008)}]{ForsterSchawinski2008}
{F{\"o}rster} F., {Schawinski} K., 2008, \mnras, 388, L74

\bibitem[{{Gezari} {et~al}\mbox{.}(2012){Gezari}, {Chornock}, {Rest}, {Huber},
  {Forster}, {Berger}, {Challis}, {Neill}, {Martin}, {Heckman}, \&
  {Lawrence}}]{Gezari2012}
{Gezari} S. {et~al.}, 2012, \nat, 485, 217

\bibitem[{{Graham}(2001)}]{Graham2001}
{Graham} A.~W., 2001, \aj, 121, 820

\bibitem[{{H{\"a}ring} \& {Rix}(2004)}]{Haring2004}
{H{\"a}ring} N., {Rix} H.-W., 2004, \apjl, 604, L89

\bibitem[{{Harrison}(2011)}]{Harrison2011}
{Harrison} D.~L., 2011, Experimental Astronomy, 31, 157

\bibitem[{{Hatano}, {Branch} \& {Deaton}(1998){Hatano}, {Branch}, \&
  {Deaton}}]{Hatano1998}
{Hatano} K., {Branch} D., {Deaton} J., 1998, \apj, 502, 177

\bibitem[{{Herrero-Illana}, {P{\'e}rez-Torres} \&
  {Alberdi}(2012){Herrero-Illana}, {P{\'e}rez-Torres}, \&
  {Alberdi}}]{Herrero-Illana2012}
{Herrero-Illana} R., {P{\'e}rez-Torres} M.~{\'A}., {Alberdi} A., 2012, \aap,
  540, L5

\bibitem[{{Holoien} {et~al}\mbox{.}(2014){Holoien}, {Prieto}, {Bersier},
  {Kochanek}, {Stanek}, {Shappee}, {Grupe}, {Basu}, {Beacom}, {Brimacombe},
  {Brown}, {Davis}, {Jencson}, {Pojmanski}, \& {Szczygie{\l}}}]{Holoien2014}
{Holoien} T.~W.-S. {et~al.}, 2014, \mnras, 445, 3263

\bibitem[{{Hsiao} {et~al}\mbox{.}(2007){Hsiao}, {Conley}, {Howell}, {Sullivan},
  {Pritchet}, {Carlberg}, {Nugent}, \& {Phillips}}]{Hsiao2007}
{Hsiao} E.~Y., {Conley} A., {Howell} D.~A., {Sullivan} M., {Pritchet} C.~J.,
  {Carlberg} R.~G., {Nugent} P.~E., {Phillips} M.~M., 2007, \apj, 663, 1187

\bibitem[{{James} \& {Anderson}(2006)}]{JamesAnderson2006}
{James} P.~A., {Anderson} J.~P., 2006, \aap, 453, 57

\bibitem[{{Jordi} {et~al}\mbox{.}(2010){Jordi}, {Gebran}, {Carrasco}, {de
  Bruijne}, {Voss}, {Fabricius}, {Knude}, {Vallenari}, {Kohley}, \&
  {Mora}}]{Jordi2010}
{Jordi} C. {et~al.}, 2010, \aap, 523, A48

\bibitem[{{Kankare} {et~al}\mbox{.}(2012){Kankare}, {Mattila}, {Ryder},
  {V{\"a}is{\"a}nen}, {Alberdi}, {Alonso-Herrero}, {Colina}, {Efstathiou},
  {Kotilainen}, {Melinder}, {P{\'e}rez-Torres}, {Romero-Ca{\~n}izales}, \&
  {Takalo}}]{Kankare2012}
{Kankare} E. {et~al.}, 2012, \apjl, 744, L19

\bibitem[{{Kesden}(2012)}]{Kesden2012}
{Kesden} M., 2012, \prd, 85, 024037

\bibitem[{{Krone-Martins} {et~al}\mbox{.}(2013){Krone-Martins}, {Ducourant},
  {Teixeira}, {Galluccio}, {Gavras}, {dos Anjos}, {de Souza}, {Machado}, \& {Le
  Campion}}]{Krone-Martins2013}
{Krone-Martins} A. {et~al.}, 2013, \aap, 556, A102

\bibitem[{{Lackner} \& {Gunn}(2012)}]{LacknerGunn2012}
{Lackner} C.~N., {Gunn} J.~E., 2012, \mnras, 421, 2277

\bibitem[{{Li} {et~al}\mbox{.}(2011{\natexlab{a}}){Li}, {Chornock}, {Leaman},
  {Filippenko}, {Poznanski}, {Wang}, {Ganeshalingam}, \& {Mannucci}}]{Li2011c}
{Li} W., {Chornock} R., {Leaman} J., {Filippenko} A.~V., {Poznanski} D., {Wang}
  X., {Ganeshalingam} M., {Mannucci} F., 2011{\natexlab{a}}, \mnras, 412, 1473

\bibitem[{{Li} {et~al}\mbox{.}(2011{\natexlab{b}}){Li}, {Leaman}, {Chornock},
  {Filippenko}, {Poznanski}, {Ganeshalingam}, {Wang}, {Modjaz}, {Jha}, {Foley},
  \& {Smith}}]{Li2011}
{Li} W. {et~al.}, 2011{\natexlab{b}}, \mnras, 412, 1441

\bibitem[{{Lodato} \& {Rossi}(2011)}]{LodatoRossi2011}
{Lodato} G., {Rossi} E.~M., 2011, \mnras, 410, 359

\bibitem[{{MacArthur}, {Courteau} \& {Holtzman}(2003){MacArthur}, {Courteau},
  \& {Holtzman}}]{MacArthur2003}
{MacArthur} L.~A., {Courteau} S., {Holtzman} J.~A., 2003, \apj, 582, 689

\bibitem[{{MacLeod} {et~al}\mbox{.}(2012){MacLeod}, {Ivezi{\'c}}, {Sesar}, {de
  Vries}, {Kochanek}, {Kelly}, {Becker}, {Lupton}, {Hall}, {Richards},
  {Anderson}, \& {Schneider}}]{MacLeod2012}
{MacLeod} C.~L. {et~al.}, 2012, \apj, 753, 106

\bibitem[{{Magnelli} {et~al}\mbox{.}(2014){Magnelli}, {Lutz}, {Saintonge},
  {Berta}, {Santini}, {Symeonidis}, {Altieri}, {Andreani}, {Aussel},
  {B{\'e}thermin}, {Bock}, {Bongiovanni}, {Cepa}, {Cimatti}, \&
  {Conley}}]{Magnelli2014}
{Magnelli} B. {et~al.}, 2014, \aap, 561, A86

\bibitem[{{Maller} {et~al}\mbox{.}(2009){Maller}, {Berlind}, {Blanton}, \&
  {Hogg}}]{Maller2009}
{Maller} A.~H., {Berlind} A.~A., {Blanton} M.~R., {Hogg} D.~W., 2009, \apj,
  691, 394

\bibitem[{{Maoz} {et~al}\mbox{.}(2011){Maoz}, {Mannucci}, {Li}, {Filippenko},
  {Della Valle}, \& {Panagia}}]{Maoz2011}
{Maoz} D., {Mannucci} F., {Li} W., {Filippenko} A.~V., {Della Valle} M.,
  {Panagia} N., 2011, \mnras, 412, 1508

\bibitem[{{Mattila} {et~al}\mbox{.}(2012){Mattila}, {Dahlen}, {Efstathiou},
  {Kankare}, {Melinder}, {Alonso-Herrero}, {P{\'e}rez-Torres}, {Ryder},
  {V{\"a}is{\"a}nen}, \& {{\"O}stlin}}]{Mattila2012}
{Mattila} S. {et~al.}, 2012, \apj, 756, 111

\bibitem[{{Melinder} {et~al}\mbox{.}(2012){Melinder}, {Dahlen}, {Menc{\'{\i}}a
  Trinchant}, {{\"O}stlin}, {Mattila}, {Sollerman}, {Fransson}, {Hayes},
  {Kankare}, \& {Nasoudi-Shoar}}]{Melinder2012}
{Melinder} J. {et~al.}, 2012, \aap, 545, A96

\bibitem[{{Perryman} {et~al}\mbox{.}(2001){Perryman}, {de Boer}, {Gilmore},
  {H{\o}g}, {Lattanzi}, {Lindegren}, {Luri}, {Mignard}, {Pace}, \& {de
  Zeeuw}}]{Perryman2001}
{Perryman} M.~A.~C. {et~al.}, 2001, \aap, 369, 339

\bibitem[{{Rees}(1988)}]{Rees1988}
{Rees} M.~J., 1988, \nat, 333, 523

\bibitem[{{Riello} \& {Patat}(2005)}]{RielloPatat2005}
{Riello} M., {Patat} F., 2005, \mnras, 362, 671

\bibitem[{{Schlegel}, {Finkbeiner} \& {Davis}(1998){Schlegel}, {Finkbeiner}, \&
  {Davis}}]{Schlegel1998}
{Schlegel} D.~J., {Finkbeiner} D.~P., {Davis} M., 1998, \apj, 500, 525

\bibitem[{{Shaw}(1979)}]{Shaw1979}
{Shaw} R.~L., 1979, \aap, 76, 188

\bibitem[{{Shen} {et~al}\mbox{.}(2003){Shen}, {Mo}, {White}, {Blanton},
  {Kauffmann}, {Voges}, {Brinkmann}, \& {Csabai}}]{Shen2003}
{Shen} S., {Mo} H.~J., {White} S.~D.~M., {Blanton} M.~R., {Kauffmann} G.,
  {Voges} W., {Brinkmann} J., {Csabai} I., 2003, \mnras, 343, 978

\bibitem[{{Siddiqui} {et~al}\mbox{.}(2014){Siddiqui}, {Els}, {Guerra}, {Cheek},
  {Mora}, \& {O'Mullane}}]{Siddiqui2014}
{Siddiqui} H., {Els} S.~G., {Guerra} R., {Cheek} N., {Mora} A., {O'Mullane} W.,
  2014, in Society of Photo-Optical Instrumentation Engineers (SPIE) Conference
  Series, Vol. 9149, Society of Photo-Optical Instrumentation Engineers (SPIE)
  Conference Series, p.~2

\bibitem[{{Simien} \& {de Vaucouleurs}(1986)}]{SimienVaucouleurs1986}
{Simien} F., {de Vaucouleurs} G., 1986, \apj, 302, 564

\bibitem[{{Smartt}(2009)}]{Smartt2009}
{Smartt} S.~J., 2009, \araa, 47, 63

\bibitem[{{van Leeuwen}(2007)}]{vanLeeuwen07}
{van Leeuwen} F., 2007, \aap, 474, 653

\bibitem[{{van Velzen} \& {Farrar}(2014)}]{vanVelzenFarrar2014}
{van Velzen} S., {Farrar} G.~R., 2014, \apj, 792, 53

\bibitem[{{van Velzen} {et~al}\mbox{.}(2011){van Velzen}, {Farrar}, {Gezari},
  {Morrell}, {Zaritsky}, {{\"O}stman}, {Smith}, {Gelfand}, \&
  {Drake}}]{vanVelzen2011}
{van Velzen} S. {et~al.}, 2011, \apj, 741, 73

\bibitem[{{Wyrzykowski} {et~al}\mbox{.}(2012){Wyrzykowski}, {Hodgkin},
  {Blogorodnova}, {Koposov}, \& {Burgon}}]{Wyrzykowski2012}
{Wyrzykowski} L., {Hodgkin} S., {Blogorodnova} N., {Koposov} S., {Burgon} R.,
  2012, arXiv:1210.5007

\bibitem[{{Wyrzykowski} {et~al}\mbox{.}(2014){Wyrzykowski},
  {Kostrzewa-Rutkowska}, {Koz{\l}owski}, {Udalski}, {Poleski}, {Skowron},
  {Blagorodnova}, {Kubiak}, {Szyma{\'n}ski}, {Pietrzy{\'n}ski},
  {Soszy{\'n}ski}, {Ulaczyk}, {Pietrukowicz}, \& {Mr{\'o}z}}]{Wyrzykowski2014}
{Wyrzykowski} {\L}. {et~al.}, 2014, \actaa, 64, 197

\end{thebibliography}

\bsp

\appendix

\section[]{\textsc{gibis} simulation tool} \label{sec:gibis}

The main simulation tool used in this paper is \textsc{gibis},  \gaia Instrument and Basic Image Simulator,  as described by \cite{babusiaux05} and \cite{babusiaux13}.
\textsc{gibis} is a maximum-detail pixel-level simulator for \textit{Gaia}, which given a source location and a deterministic NSL unambiguously determines the number of scans, and their orientation,  over the course of the mission. 
\textsc{gibis} can simulate both point-like and extended sources. To create a controlled configuration for studying the detection efficiency, we used the option given by  the \textsc{gibis} simulator for a user-defined layout of  sources about a given location on the sphere. This lets us simulate galaxies and galaxy point-source pairs with different properties, such as magnitude, size, colour and angular separation. 

The simulation of galaxies within \textsc{gibis} uses two different light profiles, one for the bulge and one for the disc. Bulges follow the De Vaucouleurs' \citep{deVaucouleurs1948} profile, where the intensity at radius $r$ is given by the following expression:
\begin{equation}
I_\mathrm{b}(r) = I_{\mathrm{e}} \  \text{exp} \left[ -b_n \left( \left[ \frac{r}{r_\mathrm{e}} \right]^{1/4}-1 \right)  \right]
\label{eq:vauc}
\end{equation}
where $r_\mathrm{e}$  is the effective radius containing half of the light from the galaxy, and $b_n=7.66925$ for n=4 as in the case of De Vaucouleurs' profile, and $I_\mathrm{e} = I_\mathrm{0} e^{-b_n}$, where $I_\mathrm{0}$ is the central intensity.

The disc component is described by the exponential profile, 
\begin{equation}
I_\mathrm{d}(r) = I_{\mathrm{d0}} \ \text{exp} \left( -\frac{r}{h} \right)
\label{eq:exp_prof}
\end{equation}
where $I_{\mathrm{d}0}$ is the central disc intensity and $h$ is the disc exponential scale length, which is correlated to the bulge size $r_\mathrm{e}/h=0.22 \pm 0.09$ \citep{MacArthur2003}. These two profiles are combined using the appropriate weights according to the galaxy bulge-to-total $B/T$ ratio parameter.
The other parameters which complete the information required to simulate a galaxy are the $V-I$ colour,  the ellipticity given by the semi-major axial ratio $b/a$,  galaxy position angle and redshift.

\textsc{gibis} allows us to simulate the on-board detection process and window allocation. One of the \textsc{gibis} running modes is meant to reproduce the VPA on-board of the satellite. This mode is used for the source detection stage. The  VPA module performs the detection and confirmation of the source candidates seen by the satellite. As described in Section \ref{sec:gaia_data}, the source is confirmed if it is seen in the next CCD strip, AF1; this step is performed in order to discard cosmic rays and other artefacts. The algorithm is optimised for the detection of point sources, so we should expect a reduced performance in terms of detection for galaxies with respect to stellar sources. The VPA also provides an estimated \gaia $G$-magnitude for the detection, and this is used to prioritise the observations in crowded fields. Brighter objects will generally have preference (see \cite{DeBruijne2015} for details about the selection processes). The \gaia $G$-band is a broad-band covering the wavelength range from about 330 to 1050 nm \citep{Jordi2010}. In crowded regions this may determine whether a window is assigned to the source or not, and in all regions this magnitude  is used to determine the transmission priority  of the window back to Earth.\\

The output from the VPA prototype for the on-board detection is provided for each simulated transit. This consists of a list of confirmed detections and rejected spurious detections. The pixel coordinates for each detection are given and, for faint detections ($G$-magnitude $\gtrsim$ 12), sub-pixel positioning is provided as well at the resolution of 1/64th of a pixel. \textsc{gibis} makes no attempt to associate the VPA output with the individual sources it simulates, we therefore had to write our own code to cross-match the VPA output with the sources simulated by \textsc{gibis}. We know the pixel coordinates at which a given source is simulated by \textsc{gibis}. It is therefore appropriate to use a matching radius given in terms of the number of pixels  rather than directly use the angular separation between the simulated source and the detection. The sources which have a VPA detection within a radius of $\sqrt{2}$\,pixels (diagonal of a pixel) are considered to have been detected. The offsets between the simulated position and the detected position in the AL and AC directions are recorded, as  well as the VPA estimate for the $G$-magnitude of the source. Also recorded is the number of detections within a radius of $\sqrt{2}$\,pixels, which is useful in the cases where we have simulated a point source offset from a galaxy. In the case of multiple sources within the match radius, the detection is always associated to the closest simulated source.
 
 \section[]{Models and simulation details}
 
\subsection{Galaxy models} \label{sec:galmod}

The first step in creating a mock galaxy catalogue was to use the \cite{Baldry2004} $r$-band luminosity function for red and blue galaxies to populate a representative sample in the redshift range $z=(0, 0.25)$ and magnitude range $M_r = (-16, -24)$. Each galaxy received a $u-r$ colour according to the relations provided in \cite{Baldry2004}. $K$-corrections for each galaxy were computed using the \textsc{K-corrections calculator} \citep{Chilingarian2010, Chilingarian2012}. Next we assigned a $B/T$ ratio for each galaxy from the measurements of \cite{LacknerGunn2012}. The size for each bulge was computed using the absolute magnitude-size relation from \cite{Shen2003}. Each galaxy has been assigned a black hole with mass $M_{\rm BH}$, which we obtained from \cite{Haring2004}, assuming the bulge luminosity, $B/T$ and a fixed mass-to-light ratio of $\Upsilon = 4.2$ in $R$ band. Through this work we adopted the following cosmological parameters: $H_0$=70 km s$^{-1}$Mpc$^{-1}$, $\Omega_\mathrm{m}$ = 0.3 and $\Omega_{\Lambda}$ = 0.7.

\subsection{Galaxy simulation with \textsc{GIBIS}} \label{sec:galaxy_simulation_conf}

Host galaxy bulges are described using two main parameters: bulge apparent magnitude, $m_{\rm G}$, expressed in \gaia $G$-magnitudes, and the galaxy bulge angular size, characterized by the bulge effective radius, $r_\mathrm{e}$. Both of them depend on the redshift of the galaxy. In \textsc{gibis}, we simulate a grid of galaxies with a combination of sizes, from 0.1 to 4 arcsec and magnitudes, from 16 to 20 mag in steps of 0.25 of a magnitude.

The simulations were centred on Galactic coordinates which under the NSL are covered 83 times over the course of the 5 yr mission.  This number is representative of the mission sky average of 70 scans  \citep{DeBruijne2012}, and therefore these coordinates were adopted as fixed parameters in all the \textsc{gibis} simulations.

The remaining parameters are the bulge-to-total light ratio $B/T$, $V-I$ colour and the ellipticity expressed by $b/a$. For simplicity we decided to fix these parameters to the following values: $B/T$=1, $V-I$=0, $b/a$=1, which represent spherical bulges. We chose the value $V-I$=0, because the \textsc{gibis} simulator requires the $V$ magnitude as an input, and it transforms it to \gaia $G$-magnitude using the $V-I$ colour using the polynomial relation described in \cite{Jordi2010}. By setting it to zero we minimize the transformation between the two filters, which now only differ by a small offset.

\subsection{Transient models}\label{sec:tranmod}

Following the relation found by \cite{SimienVaucouleurs1986}, we used the $B/T$ values to assign our galaxies a specific Hubble class, which is needed to obtain the galaxy-specific SN rate. We use the SNuB values and relations provided by \cite{Li2011c} to simulate SNe Ia, SNe Ibc and SNe II. SNuB is the rate of SN per galaxy luminosity in the $B$ band. SNuB has units of one SN per 100 yr per 10$^{10}$ L${\odot}$($B$). We select the rates in $B$ band, rather than $K$ band (SNuK) or stellar mass (SNuM), in order to be in agreement with the galaxy catalogue, which is simulated in the optical.

SN optical signature are simulated using tabulated light curves provided in \cite{Li2011}.These are scaled according to the absolute magnitude distribution described in the same paper. $K$-corrections were applied to the light curves to account for difference in filters: from $V$ band to $G$ band, and redshift. The $K$-corrections were derived using interpolated \cite{Hsiao2007} templates for SN Ia and Peter Nugent's templates\footnote{\href{https://c3.lbl.gov/nugent/nugent\_templates.html}{https://c3.lbl.gov/nugent/nugent\_templates.html}} for SN Ibc and SN II. However, being low-redshift events ($z<$ 0.2) the effect of $K$-corrections is small. For each type, the absolute magnitude at peak is drawn from a normal distribution defined by the mean and standard deviation provided in \citep{Li2011}.  These luminosity functions are not corrected for extinction in the host galaxy and therefore represent the observed peak absolute magnitudes. By assuming these values, we assume the effects of host galaxy extinction to be similar as in the study of \cite{Li2011}. Highly reddened SNe ($A_v\geq$3) are very difficult to detect for any optical transient survey. Therefore, we would expect such objects to be systematically missed by both The Lick Observatory Supernova Search and \gaia.

The evolution of the optical signature of a TDE is due to emission from the hot disc caused by the disruption and accretion of a star next to the black hole. Generally, the accretion rate is governed by the galaxy black hole mass $M_{\rm BH}$, the disrupted star mass, $M_{\star}$, radius, $M_{\star}$ and density profile, as well as the radius of disruption or tidal radius $r_t=R_{\star} (M_{\rm BH}/M_{\rm \star})^{1/3}$ and the penetration factor $\beta=r_t/r_p$, which is the ratio of the tidal radius to the star pericentre distance. Disruptions closer to the SMBH will have a higher $\beta$ factor. 
Observationally, TDE are often considered as black bodies at temperatures generally higher than observed for SNe: from 2 $\times 10^4$ to 5 $\times 10^4$ K \citep{vanVelzen2011, Gezari2012, Arcavi2014, Holoien2014}. Because of this the $K$-corrections and the shape of the band-pass play a very important role. In this work, we used the \gaia $G$-band with effective wavelength of 600 nm to generate the light curves.

The rate of optical tidal disruption flares has been measured by \cite{vanVelzenFarrar2014}. Besides a large Poisson uncertainty due to the low number of  TDEs that have been found in the systematic survey, an accurate measurement of the rate of optical tidal disruption flares is limited by the uncertainty of the light curve. This systematic uncertainty implies that the derived disruption rate depends on the adopted light curve model. In our study, we reduce this systematic uncertainty by using the same light curves to simulate the optical evolution of TDE as the ones that were used by \cite{vanVelzenFarrar2014} to derive the disruption rate. These light curves are divided into two main groups, driven by observations and driven by models. 

\begin{itemize}
\item  The observed light curves come from reported Pan-STARRS TDE candidate events (PS1 from now on): PS1-10jh \citep{Gezari2012} and PS1-11af \citep{Chornock2014}. We select either the PS1-10jh or PS1-11af light curve according to the black hole mass of the host galaxy. For example, for $M_{\rm BH} < 10^{6.6}\ \mathrm{M}{\sun}$ we choose PS1-10jh, for $M_{\rm BH} > 10^{6.9}\ \mathrm{M}{\sun}$, we choose PS1-11af. For masses in between these two values, the probability of selecting PS1-11af increases linearly with mass between the two masses. These light curves are fixed and their shape does not depend on the adopted black hole mass. The TDE rate derived from these light curves is $2.0 \times 10^{-5}$galaxy$^{-1}$yr$^{-1}$.

 \item The theoretical light curves are given by the model from \cite{LodatoRossi2011} (LR11 from now on). This model describe the monochromatic TDE light curves as emission from two main components: a geometrically thin and optically thick disc, and a radiative wind, coming from the fraction of matter that is expelled during super-Eddington accretion regime. Similarly to the method described in \cite{vanVelzenFarrar2014}, we normalize the light curve luminosity to two known events: TDE1 or TDE2 from \cite{vanVelzen2011}. We select the event according to the mass of the host galaxy using the same approach as for observed light curves. The TDE rate used for this light curves is $1.7 \times 10^{-5}$galaxy$^{-1}$yr$^{-1}$. In this case, the light curve characteristics depend on the black hole mass.

To make the model more tractable, we have made a few simplifying assumptions. We fixed the disruption to be of a main sequence star, $M_{\star}=1 \mathrm{M}{\sun}$, radius $R_{\star}=1 R{\odot}$, having a penetration factor $\beta=1$.

\end{itemize}

\subsection{Transient simulation with \textsc{GIBIS}} \label{sec:tran_simulation_conf}

The simulation of transients close to their host galaxies was done using four main parameters:
 $m_{\rm G}$ and $r_\mathrm{e}$ for the galaxy, and for the transient its apparent magnitude $m_{\rm T}$ and angular separation to the centre of its host galaxy, $\theta$. 

The configuration for these four parameters in our \textsc{gibis} simulations was selected as follows. Transient magnitudes were simulated from 15 to 20.4, in steps of 0.2 mag, so that we go slightly beyond the 20th limiting magnitude for \gaia. A non-linear progression between 0 and 1 arcsec, was used for the angular separations, with values given by $\theta= (0, 0.05, 0.1, 0.15,  0.2, 0.3, 0.4, 0.5, 0.7, 1)$. We created a different simulation for each galaxy magnitude $m_{\rm G}=16 - 20$ mag in steps of 1 magnitude and effective radius $r_\mathrm{e}=(0.5, 1, 1.5, 2.5, 3.5, 4.5)$ arcsec. This range of bulge sizes is appropriate for compact galaxies at redshifts larger than 0.01. According to \cite{deSouza2014}, very nearby galaxies with apparent angular sizes $r_\mathrm{e}>$4.72 arcsec will not be detected, as their detection will be limited by the maximum size of the window transmitted to Earth. Therefore, bulges larger than 4.5 arcsec are not included in the simulation.

\end{document}